\documentclass[a4paper,fleqn]{cas-sc}

\usepackage[authoryear]{natbib}
\usepackage{amsmath}
\usepackage{graphicx}
\usepackage{subcaption}
\usepackage{float}
\usepackage{placeins}
\usepackage{lipsum}
\usepackage{soul,color}
\usepackage{subcaption}
\usepackage{pdflscape}
\usepackage{rotating}
\usepackage{makecell}

\numberwithin{equation}{section}

\usepackage[pagewise]{lineno}

\def\tsc#1{\csdef{#1}{\textsc{\lowercase{#1}}\xspace}}
\tsc{WGM}
\tsc{QE}
\tsc{EP}
\tsc{PMS}
\tsc{BEC}
\tsc{DE}

\begin{document}
\let\WriteBookmarks\relax
\def\floatpagepagefraction{1}
\def\textpagefraction{.001}
\shorttitle{Optimal Parking Planning for Shared Autonomous Vehicles}
\shortauthors{S. Choi and J. Lee}

\title [mode = title]{Optimal Parking Planning for Shared Autonomous Vehicles}




\author[1]{Seongjin Choi}[orcid=0000-0001-7140-537X]
\ead{seongjin.choi@mcgill.ca}
\credit{Conceptualization of this study, Methodology, Software, Validation, Formal analysis,  Writing - original draft}

\author[2]{Jinwoo Lee}[orcid=0000-0002-6692-9715]
\cormark[1]
\ead{lee.jinwoo@kaist.ac.kr}
\credit{Conceptualization of this study, Methodology, Formal analysis, Writing - review and editing}

\address[1]{Department of Civil Engineering, McGill University, 817 Sherbrooke Street West, Montreal, Quebec H3A 0C3, Canada}
\address[2]{Cho Chun Shik Graduate School of Mobility, Korea Advanced Institute of Science and Technology, 193, Munji-ro, Yuseong-gu, Daejeon 34051, Republic of Korea}

\cortext[cor1]{Corresponding author}










\begin{abstract}
Parking is a crucial element of the driving experience in urban transportation systems. Especially in the coming era of Shared Autonomous Vehicles (SAVs), parking operations in urban transportation networks will inevitably change. Parking stations will serve as storage places for unused vehicles and depots that control the level-of-service of SAVs. This study presents an Analytical Parking Planning Model (APPM) for the SAV environment to provide broader insights into parking planning decisions. Two specific planning scenarios are considered for the APPM: (i) Single-zone APPM (S-APPM), which considers the target area as a single homogeneous zone, and (ii) Two-zone APPM (T-APPM), which considers the target area as two different zones, such as city center and suburban area. S-APPM offers a closed-form solution to find the optimal density of parking stations and parking spaces and the optimal number of SAV fleets, which is beneficial for understanding the explicit relationship between planning decisions and the given environments, including demand density and cost factors. In addition, to incorporate different macroscopic characteristics across two zones, T-APPM accounts for inter- and intra-zonal passenger trips and the relocation of vehicles. We conduct a case study to demonstrate the proposed method with the actual data collected in Seoul Metropolitan Area, South Korea. Sensitivity analyses with respect to cost factors are performed to provide decision-makers with further insights. Also, we find that the optimal densities of parking stations and spaces in the target area are much lower than the current situations. 

\end{abstract}



\begin{keywords}
Parking\sep 
Shared Autonomous Vehicles\sep 
Optimization\sep
Planning\sep
Relocation\sep
\end{keywords}


\maketitle

\section{Introduction}
Parking is one of the crucial elements of the driving experience in urban transportation systems. However, as the number of vehicles in large cities around the globe increases rapidly, the lack of parking spaces has become a severe problem. From the perspective of the transportation system operator, the need to store vehicles is progressively increasing, which has led to the transformation of valuable real estate into parking garages \citep{nourinejad2018designing}. Moreover, from the perspective of individual users, it is required to spend more time on the road searching for empty parking spaces, which eventually worsens overall traffic conditions \citep{lam2006modeling}. As a result, to improve the overall efficiency of the transportation system, it is necessary to study efficient parking operations.

A typical vehicle spends around 95\% of its lifetime sitting in a parking space \citep{bates2012spaced}. If we can utilize these unused vehicles to serve other travel demands, there is a possibility of reducing overall system costs, including vehicle ownership and parking operation. The idea behind ‘Shared Autonomous Vehicles’ corresponds to this possibility \citep{shaheen2016mobility}. Shared Autonomous Vehicles (SAVs) are the combination of growing shared mobility services and emerging autonomous vehicle technology, which can enable cost savings, provide convenience to users, and lead to sustainable transportation by reducing vehicle usage \citep{narayanan2020shared, ko2021survey}.

In the coming era of SAVs, it is inevitable to change parking operations in urban transportation networks \citep{zhang2017parking,golbabaei2021role}. Parking stations will serve as storage places for unused vehicles and depots that control the level-of-service of SAVs. As the market penetration of SAVs increases, there will be two main changes related to parking operations. First, parking demand will be reduced because overall vehicle usage will decrease \citep{zhang2017parking,narayanan2020shared}. According to previous research, more than 25\% of individuals are willing to give up their vehicle ownership given the availability of an SAV alternative \citep{menon2019shared}. Also, one SAV can replace from 1.93 to potentially ten conventional individually owned vehicles \citep{lokhandwala2018dynamic, fagnant2014travel}. Second, the spatial distribution of parking stations (parking lots) will be fundamentally rearranged \citep{zhang2017interaction}. It will be possible to relocate parking stations outside the city center; SAVs will travel from parking stations to passengers because SAVs can travel at a low cost \citep{kroger2017autonomous}.




There have been numerous studies related to parking operations with SAVs. \cite{zhang2015exploring} used an agent-based simulation model in a hypothetical city to estimate the potential impact of an SAV system on urban parking demand. \cite{zhang2017parking} used real-world data to develop a discrete event simulation model to examine the effects of SAVs on urban parking land use. \cite{azevedo2016microsimulation} used SimMobility to analyze the impact of demand and supply of autonomous mobility on demand (AMoD) in Singapore. This study used an optimization algorithm to solve the facility location problem to find optimal locations for a fixed number of parking stations. In \cite{kondor2018estimating}, the distance an SAV can travel to the nearest parking station is considered a constraint in estimating the required number of parking stations. \cite{okeke2020impacts} presented a case study at the University of West England to analyze the impacts of autonomous vehicle technology on parking operations. \cite{okeke2020impacts} used agent-based simulation and a parking model and concluded that SAVs would allow parking stations to be relocated outside city centers.



Despite many studies on parking operations with the dominance of SAVs, most studies have used simulation-based approaches. However, such approaches have limitations. They require massive effort and cost (i) to acquire appropriate and detailed data and (ii) to set up and run the computationally heavy simulations to obtain appropriate solutions. Consequently, solutions from simulation-based approaches rely highly on the quality of collected data so that proper solutions can only be found when there exists a sufficient amount of explanatory data for simulation. Moreover, simulation results for a particular situation are difficult to generalize and apply to different situations. The findings from the simulation results from a certain city cannot be directly applied to another city's planning problem, since the results are `site-specific.'

In this study, to overcome the limitations of simulation-based approaches, we present an Analytical Parking Planning Model (APPM) with SAVs. Usually, analytical models focus primarily on functional systemic relationships between planning variables and the objective function. Thanks to the mathematical approximations in their parsimonious modeling, only macroscopic data of target cities (or areas) are needed instead of the detailed data required in simulation analysis \citep{daganzo2019public}. Analytical models are used to identify generally applicable insight on planning decisions; if detailed data are available, this insight can be further improved and refined through simulations for target areas if necessary. Similarly, the proposed model in this study explicitly explains the inter-relationship among the parking planning variables together with the other important exogenous factors, such as land cost and vehicle operating costs, in a closed form. The required data to use the proposed model is much simpler than the data required for simulation-based approaches, which overcomes the first limitation of simulation-based approaches. Also, using a closed form can overcome the significant computation inefficiency of simulation-based approaches, since the solutions can be calculated by one shot. 





We use two scenarios to describe parking operations in a given urban traffic network, the \textit{Single-zone Analytical Parking Planning Model} (S-APPM) and \textit{Two-zone Analytical Parking Planning Model} (T-APPM) to properly design a model structure that describes the parking operation with SAVs. S-APPM considers the target region a single zone with macroscopic characteristics for parking planning decisions. As a result, the derived optimal parking planning decisions are assumed to be the same in all sub-regions in the target region. On the other hand, T-APPM considers the target region as two distinguishable zones, usually represented as a city center and suburb. As a result, it is possible to consider the effects of different macroscopic characteristics of each zone (such as passenger demand, land cost, average speed, etc.) on parking planning decisions. Also, we conduct case studies for each model to demonstrate the sensitivity of the model on the cost factors and the effect of relocation between zones on parking planning.
The contributions in this paper are summarized as follows:
\begin{itemize}
    \item To the best of our knowledge, it is the first to propose analytical models for parking planning, especially with SAVs. 
    \item We formulate the parking operation problems with total operating cost as an objective function, and we carefully derive the parking operation variables in the objective functions for S-APPM (Section \ref{SAPPM}) and T-APPM (Section \ref{sec:tappm}). The models explicitly explain the inter-relationship among the variables together with the other important exogenous factors.
    \item We conduct the case studies to give general insights on the proposed model with extensive sensitivity analyses on cost parameters for each parking operation variable. (Section \ref{sec:case_study_sappm}, Section \ref{sec:case_study_tappm})
\end{itemize}

The organization of this paper is as follows. We present S-APPM in Section \ref{SAPPM} and solve the objective function to derive the decision variables and parking operational variables. Then, in Section \ref{sec:case_study_sappm}, a case study of Seoul, South Korea, is presented to further elaborate on our findings for S-APPM. T-APPM is presented in Section \ref{sec:tappm} by extending S-APPM with the relocation of vehicles between two zones. A case study of the Seoul Metropolitan Area, including Seoul and other cities near Seoul, is presented in Section \ref{sec:case_study_tappm}. Finally, in Section \ref{sec:conclusion}, we conclude our study by summarizing the key findings and contribution of this study, as well as addressing limitations and proposing future research. The notations used in this paper are listed in Appendix \ref{tab:notations}.


\section{Single-Zone Analytical Parking Planning Model}\label{SAPPM}
\begin{figure}[!t]
  \centering
  \includegraphics[width=0.7\textwidth]{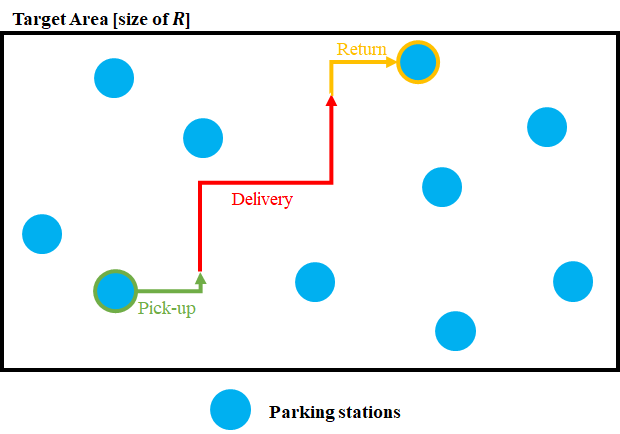}
  \caption{Operational scenario of parking stations with Shared Autonomous Vehicles. Blue circles represent parking stations. SAV moves from one parking station to the origin location of the passenger (green), picks up the passenger, and delivers the passenger to the destination (red); SAV then moves to the nearest parking station (yellow).}\label{fig:scenario}
\end{figure}


To be consistent with the previous demand-responsive shared mobility model without the parking state \citep{daganzo2019general}, we consider a given target region with a size of $R$ [$\mathrm{km^2}$] simplified into a single zone with uniform origin and destination (O-D) demands $\lambda_t$ [$\mathrm{veh/km^2/hr}$], which vary by time window $t$. Time windows associated with the daily maximum and minimum demands are designated by $t_{max}$ and $t_{min}$, respectively. Figure \ref{fig:scenario} shows the operation scenario for parking stations with SAVs. The blue circles represent the parking stations installed in the target region. The density of the parking stations is denoted as $x$  $\mathrm{[stations/km^2]}$, parking spaces in the target region is denoted as $y$ $\mathrm{[spaces/km^2]}$, and the average parking spaces per station $z$ $\mathrm{[spaces/station]}$ is $y/x$. The red line represents a single passenger demand in the scenario. When a passenger demand is generated, an SAV in the nearest not-empty parking station is assigned to the passenger, cruises to the origin location of the passenger, and picks up the passenger. After pickup, the SAV delivers the passenger to the destination. Then, the SAV cruises to the nearest not-full parking station and awaits the next passenger assignment.

\subsection{Optimization Framework}\label{sec:sappm_objective}
The objective of this model is to minimize total operating cost in the target region by determining three variables related to parking space and SAV fleet planning:
\begin{itemize}    \itemsep-0.5em
    \item $x$ -- density of parking stations $[\mathrm{stations/km^2}]$
    \item $y$ -- density of parking spaces $[\mathrm{spaces/km^2}]$
    \item $m$ -- number of SAV fleets $[\mathrm{veh}]$
\end{itemize}
The objective function of the parking operation consists of three different operation costs: (i) parking station costs, (ii) parking space costs, and (iii) fleet costs. The objective function $J$ of the parking operation  [$\mathrm{\$/day}$] is to minimize the overall daily average operation cost ($Cost$), formulated as a function of the planning variables, $x$, $y$, and $m$, as follows:

\begin{equation}\label{eq:ob}
  \begin{split}
  J = \min_{x,y,m} Cost(x,y,m) = \min_{x,y,m} {\Big(C_x xR + C_y yR + C_m m \Big)}
  \end{split},
\end{equation}

\noindent
where $C_x$ refers to the daily average operation cost of each parking station, such as the rental cost or prorated purchase cost for the land and built infrastructure on it, except for the variable costs depending on the number of parking spaces at the station $\mathrm{[\$/stations/day]}$; $C_y$ stands for the daily operation cost of a unit parking space, except for the cost components included in $C_x$ $\mathrm{[\$/spaces/day]}$; and $C_m$ indicates the daily operation cost of each vehicle $[\mathrm{\$/veh/day}]$.


For passenger convenience, measures of Level-Of-Service (LOS) of SAV operation related to passenger waiting time can be used as constraints. In demand-responsive mobility services, ensuring short assigning time, i.e., the elapsed time between a call and vehicle assignment, and assigned time, i.e., the waiting time between the assignment and passenger pickup, is important \citep{lees2016minimising, daganzo2019general}. In SAV operations, as an assignment can be automatically conducted in a top-down manner from the control center, assigning time can be ignored. Thus, we take into account a LOS constraint that restricts only the average assigned time so that it does not exceed a pre-selected threshold, as in:


\begin{equation}\label{eq:ob_con}
  \begin{split}
  T_{A,t} \leq T_0, \forall t
  \end{split}.
\end{equation}
\noindent
where $T_{A,t}$ is the average passenger waiting time, only consisting of assigned time, in time window indexed by $t$ (in state $A$), that means ``Assigned'', and $T_0$ is the threshold, the maximum allowed average passenger waiting time. This constraint is likely to be binding when travel demand is the highest in $t_{max}$.

Last, it is impossible to park additional SAV more than $z=x/y$ at one parking station. This constraint is especially important when travel demand is the lowest in $t_{min}$.



\subsection{SAV Operation Model with Parking}\label{section:2.1.2}

\begin{figure}[!t]
  \centering
  \includegraphics[width=0.7\textwidth]{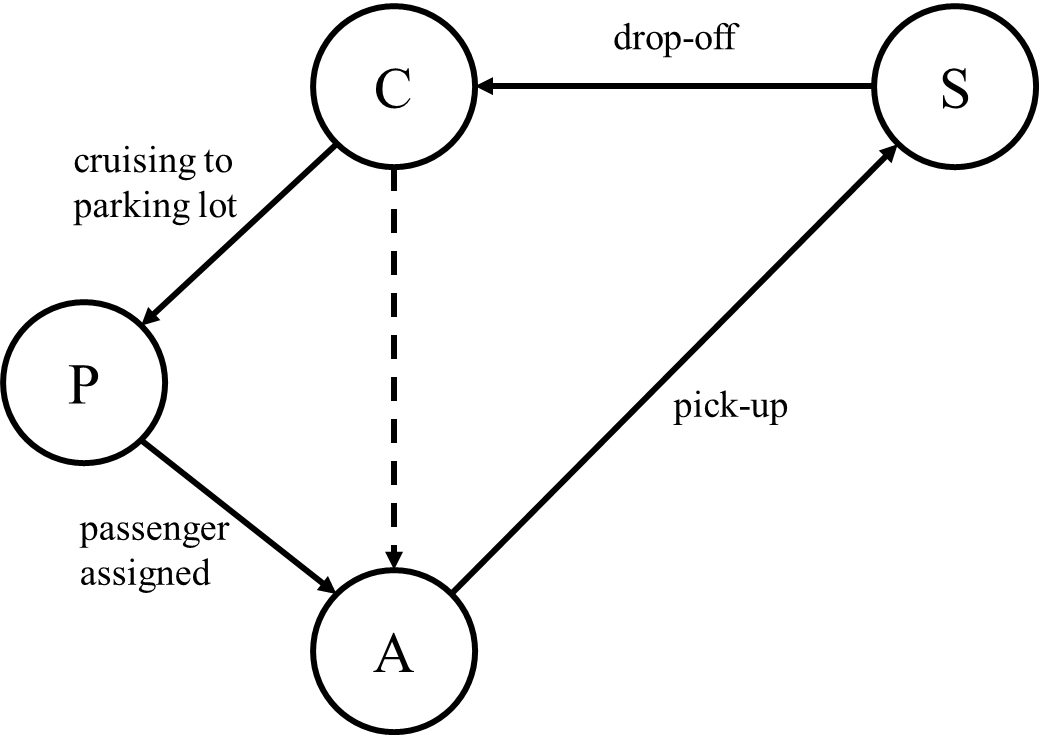}
  \caption{Workload transition network representation of Single-zone Parking Planning Model (S-APPM). Circles represent the states, and the directed lines represent transitions between states. The dotted line (from $C$ to $A$) is also possible in reality.}\label{fig:modelstate}
\end{figure}

To analytically formulate the LOS constraint as a function of the decision variables , we model SAV operations with parking planning as a workload transition network graph presented in Figure \ref{fig:modelstate}, inspired by the demand-responsive transit model proposed by \cite{daganzo2019general}. The nodes of the graph in Figure \ref{fig:modelstate} represent operational states of SAVs, and the links stand for the transitions between them. An SAV is in either of four different states:


\begin{itemize}    \itemsep-0.5em
    \item P (Parked): the vehicle is idle and parked at the parking station;
    \item A (Assigned): once assigned to a passenger, the vehicle is moving to the passenger’s current location from the previous parking station;
    \item S (Serving): the vehicle is delivering a passenger from his/her origin to destination;
    \item C (Cruising to return): After serving, the empty vehicle is cruising to the nearest not-full parking station. 
\end{itemize}

Once a passenger engages service, we assume that a vehicle parked at the nearest non-empty station is assigned to the passenger, resulting in a change of operational state of the assigned vehicle from state $P$ to state $A$ ($P \rightarrow A$). Then, the assigned vehicle moves from the parking station to the origin of the passenger’s trip ($A \rightarrow S$). When the vehicle arrives at the origin of the passenger’s trip, the vehicle picks up the passenger and travels from origin to destination. The operational state changes from \textit{Serving} $S$ to \textit{Cruising to return} $C$ when the passenger gets off the vehicle ($S \rightarrow C$). After completion of the passenger’s trip, the vehicle moves to the nearest not-full parking station and waits for the next assignment ($C \rightarrow P$).


In reality, it is also possible to allocate a new passenger request to the nearest vehicle in state $C$. In other words, vehicles that are moving to a parking station can be allocated to another passenger’s request before they arrive at the parking station, as represented by the dotted line in Figure \ref{fig:modelstate}. However, for the mathematical simplicity of this model, we neglect this state transition. Nonetheless, the proposed model can provide a lower bound on the efficiency of the system because including the state transition from state $C$ to state $A$ will increase the overall efficiency of the SAV system through extra operational flexibility. The lower bound can provide us with information about the minimum benefits of SAV operations with optimal parking and fleet planning, which is particularly important to determine the feasibility of introducing the system.

\begin{figure}[!t]
  \centering
  \includegraphics[width=\textwidth]{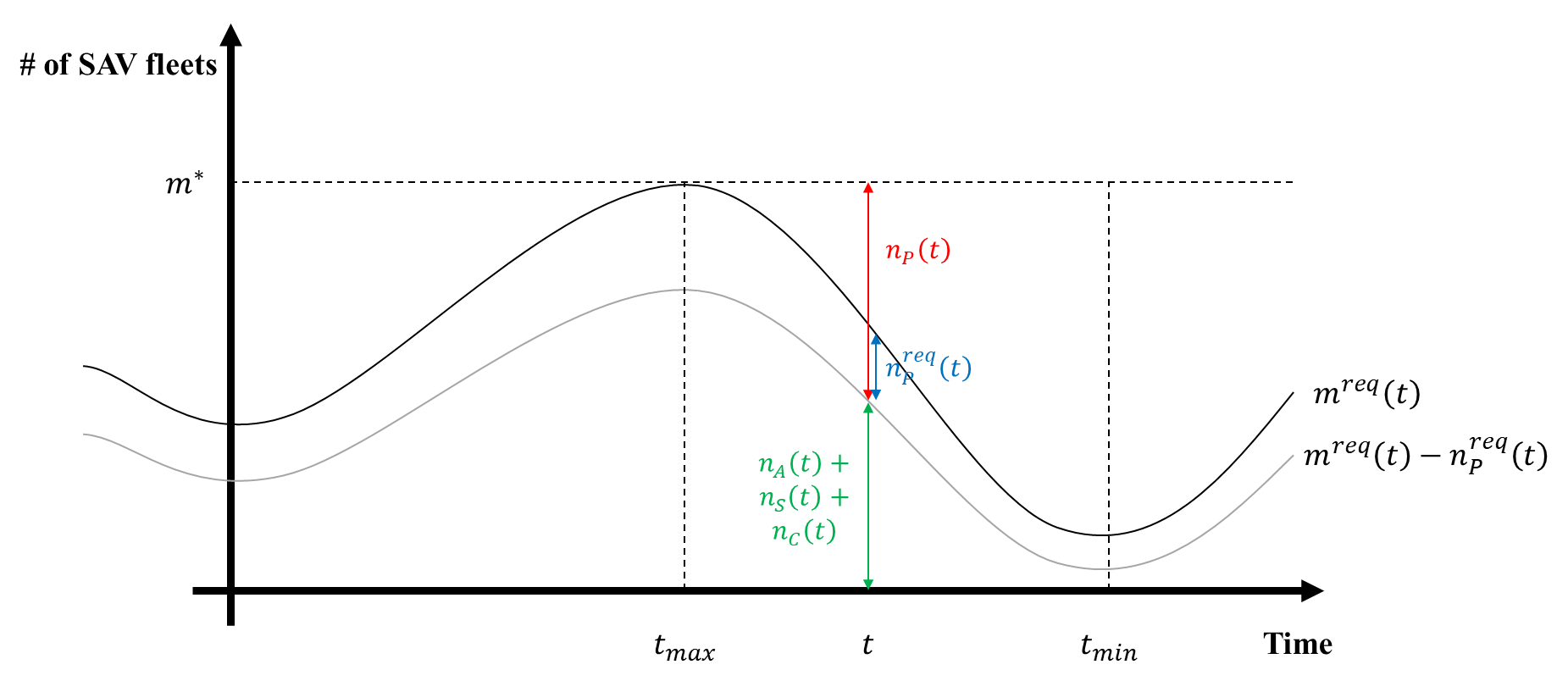}
  \caption{Graphical expression of change of the number of SAV fleets over time. $t_{max}$ represents the time window associated with maximum demand and $t_{min}$ represents the time window associated with minimum demand. At a given time window $t$, the red portion represents the number of SAV fleets parked at parking stations ($n_P (t)$), the blue portion represents the required number of SAV fleets in parking station, and the green portion represents the number of SAV fleets actively running on the roads. 
  }\label{fig:numveh}
\end{figure}

The fleet size in each state in the given time window, denoted by $n_A(t)$, $n_S(t)$, $n_C(t)$, and $n_P(t)$, respectively, vary depending on the time-dependent passenger demand $\lambda_t$, as well as the decision variable. 

\begin{equation}
\label{eq:m}
    \begin{split}
        m(t) =  n_A(t) + n_S(t) + n_C(t) + n_P(t),
    \end{split}
\end{equation}

The required number of fleets in time window $t$, needed to ensure user LOS, can be defined as the summation of the required number of fleets in each state to ensure the LOS as shown in Equation \ref{eq:mreq}:


\begin{equation}
\label{eq:mreq}
    \begin{split}
        m^{req}(t) =  n_A^{req}(t) + n_S^{req}(t) + n_C^{req}(t) + n_P^{req}(t),
    \end{split}
\end{equation}

\noindent
where $n_A^{req}(t)$, $n_S^{req}(t)$, $n_C^{req}(t)$, and $n_P^{req}(t)$ refer to the required number of fleets in state $A$, $S$, $C$, and $P$ , respectively.




As shown in Figure \ref{fig:numveh}, the SAV fleet size $m$ is constant over the whole day, so the minimum required fleet size $m^*$ is equal to the maximum required number of SAVs among all the time windows of the day, i.e., the required fleet size in time window $t_{max}$ when the demand is at the maximum level, as shown in Equation \ref{eq:mopt}:

\begin{equation}
\label{eq:mopt}
    \begin{split}
    m^* = \max_t \left( m^{req}(t) \right)
    \end{split}.
\end{equation}

The three components on the right-hand side of Equation \ref{eq:mreq}, $n_A^{req}(t)$, $n_S^{req}(t)$, and $n_C^{req}(t)$, can be expressed based on Little’s Law with exogenous and endogenous planning variables. In other words, the fleet sizes in each state are calculated by multiplying the demand ($\lambda_t R$) by the expected time spent in each state ($T_{A}$, $T_{S}$ , and $T_{C}$). The last component, $n_P^{req} (t)$, is a buffer to ensure that there are not too many (not to exceed the parking space limit at a station) and not too few (to ensure the level-of-service by guaranteeing an idle vehicle to assign to any demand generated nearby) vehicles at the parking station. $n_P^{req} (t)$ can be derived based on the variance of the vehicle inflow and outflow at each parking station.


First, the fleet size in state $A$ in time window $t$, $n_A^{req} (t)$, is derived as follows:

\begin{equation}\label{eq:n_A}
    \begin{split}
    n_A^{req}(t) = \lambda_t R T_{A,t}
    \end{split},
\end{equation}

\noindent
where, as used in Equation \ref{eq:ob_con}, SAVs' assigned time $T_{A,t}$ is equal to the average passenger waiting time.

In some cases, especially when passenger demand is high, it is possible that the parking station nearest to the origin of the passenger does not have an idle SAV to serve passenger demand. In this case, it is required to send an idle SAV from the second-nearest parking station, and so on. As a result, we consider a confidence level $p$, defined as the probability that a passenger is served by an SAV from the nearest parking station. The number of fleets in state $A$ can be expressed as Equation \ref{eq:n_A_extend}. If $p$ is sufficiently high, $p\rightarrow1$, the probability that an SAV from the $i$-th nearest parking station is assigned to a passenger, $(1-p)^i\cdot p$, becomes almost zero, so we ignore the corresponding terms. 


\begin{equation}\label{eq:n_A_extend}
    \begin{split}
    n_A^{req}(t) 
    &= \lambda_t R \left(p T^1_{A,t} + (1-p)\left(p T^2_{A,t} + (1-p)(p T^3_{A,t} +(1-p)(\cdots) \right) \right)\\
    &\approx \lambda_t R \left(p T^1_{A,t} + (1-p)\left(p T^2_{A,t} \right) \right)\\
    &= \lambda_t R \left( p T^1_{A,t} + (1-p)\cdot \alpha pT_{A,t}^1 \right) \\
    &= \lambda_t R T^1_{A,t} \left( p + \alpha p - \alpha p^2 \right) \\
    \end{split},
\end{equation}
\noindent
where $T_{A,t}^i$ refers to the average travel time from the $i$-th nearest parking station to the origin of the passenger, and $\alpha$ is the incremental ratio of the travel time, i.e., $\alpha \equiv T_{A,t}^2 / T_{A,t}^1$. From Equations \ref{eq:n_A} and \ref{eq:n_A_extend}, $T_{A,t}$ is $T_{A,t}^1 (p+\alpha p - \alpha p^2 )$.

Second, the number of fleets in state $S$, $n_S^{req}(t)$, is derived as follows:
\begin{equation}\label{eq:n_S}
    \begin{split}
    n_S^{req}(t) = \lambda_t R T_{S,t} = \lambda_t R \frac{l_t}{v_t}
    \end{split},
\end{equation}

\noindent
As $T_{S,t}$ is the average travel time from origin to the destination for all passengers in time window $t$, it can be simply derived with average trip length $l_t$ $\mathrm{[km]}$ considering the circuity of roads and the average speed $v_t$ $[\mathrm{km/hr}]$.

Third, the fleet size in state $C$ in time window t is formulated as follows:
\begin{equation}\label{eq:n_C}
  \begin{split}
  n_C^{req}(t) = \lambda_t R T_{C,t}
  \end{split},
\end{equation}

\noindent
where $T_{C,t}$ is the expected time a vehicle spends in state $C$. $T_{C,t}$ is the travel time from the destination of the passenger to the nearest parking station. When passenger demand is low, it is possible that the parking station nearest the destination of the passenger will be full and there are no parking spaces left. In this case, it is required to send the SAV fleet to the second-nearest parking station, and so on. As a result, we consider a confidence level $q$, the probability that the nearest parking station is not full. The fleet size in state $C$ in time window $t$ can be expressed as follows:

\begin{equation}\label{eq:n_C_extend}
    \begin{split}
    n_C^{req}(t) 
    &= \lambda_t R \left(q T^1_C + (1-q)\left(q T^2_C + (1-q)(q T^3_C +(1-q)(\cdots) \right) \right)\\
    &\approx \lambda_t R \left(p T^1_C + (1-q)\left(q T^2_C \right) \right)\\
    &= \lambda_t R \left( q T^1_C + (1-q)\cdot \alpha qT_C^1 \right) \\
    &= \lambda_t R T^1_C \left( q + \alpha q - \alpha q^2 \right) \\
    \end{split},
\end{equation}
\noindent
where $T_{C,t}^i$ refers to the average travel time from the destination of the passenger to the $i$-th nearest parking station and $\alpha$ is the incremental ratio of travel time between $T_{C,t}^2$ and $T_{C,t}^1$, which is equivalent to $\alpha \equiv  T_{A,t}^2 / T_{A,t}^1$ in Equation \ref{eq:n_A_extend}. Based on the assumption of uniformly distributed O-D demands and parking station, and $T^1_{C,t}$ is equal to $T^1_{A,t}$


Finally, the required fleet size in state $P$, $n_P^{req} (t)$ can be derived by accounting for the variance of the number of idle vehicles parked at each parking station. Variances of vehicles coming into a parking station and vehicles going out of a parking station are $\lambda_t R H I /x$, respectively, where $H$ is the length of the time window, and $I$ is the mean-to-variance ratio of the number of vehicles parked at each parking station. Consequently, the variance of vehicle number at a parking station during a time window is  $2 \lambda_t R H I / x$. It is assumed that, by repositioning vehicles, the number of vehicles at each parking station is rebalanced at intervals of duration H. The cost of relocation is shown to be proportional to $H^(-1/2)$  but independent of the decision variables $x$, $y$, and $m$ considered in this paper (See Section 7.2.2.1 in \cite{daganzo2019public}). Thus, the relocation cost is omitted from the cost-minimizing objective function because the relocation cost is constant if $H$ is assumed to be given.

%


When passenger demand is high, the fleet size of idle vehicles at each parking station will be at a minimum because most vehicles will be on the road serving passenger demand. In such circumstances, there must be at least a certain number of parked vehicles to guarantee the level-of-service, preventing long waiting times for passengers and assignment failures. We have already set a confidence level $p$ to ensure vehicle assignment from the nearest parking station in Equation \ref{eq:n_A_extend}. Using $p$, the required parking spaces per parking station $z^{req} (t)$ is formulated as follows:


\begin{equation}
  \begin{split}
  z^{req}(t) = \Phi^{-1}(p) \sqrt{\frac{2\lambda_{t} R H I}{x}}
  \end{split},
\end{equation}

\noindent
where $\Phi$ is the standard normal distribution. The required density of parking spaces in the target region at demand $\lambda_{t}$ is as follows:

\begin{equation}
  \begin{split}
  y^{req}(t) = \frac{x \cdot z^{req}(t)}{R} = \Phi^{-1}(p) \sqrt{\frac{2\lambda_t HIx}{R}}
  \end{split},
\end{equation}

As a result, the required fleet size in state $P$ is formulated as:

\begin{equation}
  \begin{split}
  n^{req}_P(t)=y^{req}(t)R = \Phi^{-1}(p) \sqrt{2\lambda_t RHIx}
  \end{split},
  \label{eq:n_P}
\end{equation}

Finally, the density of parking station $x$ can be expressed with respect to $T^1_{A,t}$. If the parking stations are uniformly distributed, the expected value of the distance between random demand to the nearest parking station, denoted by $d(x)$, is:

\begin{equation}
  \begin{split}
  & E[d(x)] = T^1_{A,t} v\\
  & E[d(x)] \approx  \frac{\kappa}{\sqrt{x}}\\
  \end{split},
\end{equation}

\noindent
where the first expression is given by the definition of $T_{A,t}$ and the second expression is referenced from \cite{daganzo2019public}. As a result, $T_{A,t}^1$ is expressed as a function of the decision variable $x$ and vice versa:
As a result,
\begin{equation}
\label{eq:xTa}
  \begin{split}
    & T_{A,t}^1 = \frac{\kappa}{v_t \sqrt{x}}
    & x = \frac{\kappa^2}{(v_t)^2} \cdot \frac{1}{(T^1_{A,t})^2} 
  \end{split}.
\end{equation}


According to Equations \ref{eq:n_A}, \ref{eq:n_S}, \ref{eq:n_C}, and \ref{eq:n_P}, the number of fleets in each state increases as the passenger demand ($\lambda_t$) increases. 
%
%
We reasonably assume that the average ground speeds in the time windows $v_{t_{max}}$ $v_{t_{min}}$ and are the slowest and fastest of the day, i.e., $v_{t_{max}}=v_{min}$ and $v_{t_{min}}=v_{max}$, respectively. Thus, the numbers of vehicles in states $A$, $S$, $C$, and $P$ are the highest during the peak time:

\begin{equation}
\label{eq:tmax}
  \begin{split}
  & \Bigg( n_A^{req}(t_{max}) = \Bigg) \lambda_{t_{max}} \frac{\kappa R}{v_{min} \sqrt{x}} (p + \alpha p - \alpha p^2) \geq 
  \lambda_{t} \frac{\kappa R}{v_{t} \sqrt{x}}  (p + \alpha p - \alpha p^2) \Bigg( = n_A^{req}(t) \Bigg),\\
  & \Bigg( n_S^{req}(t_{max}) = \Bigg) \lambda_{t_{max}} \frac{l_{t_{max}}}{v_{min}} \geq 
  \lambda_{t} \frac{l_t}{v_{t}} \Bigg( = n_S^{req}(t) \Bigg),\\
  & \Bigg( n_C^{req}(t_{max}) = \Bigg) \lambda_{t_{max}} \frac{\kappa R}{v_{min} \sqrt{x}}  (q + \alpha q - \alpha q^2) \geq 
  \lambda_{t} \frac{\kappa R}{v_{t} \sqrt{x}}  (q + \alpha q - \alpha q^2) \Bigg( = n_C^{req}(t) \Bigg),\\  
  & \Bigg( n_P^{req}(t_{max}) = \Bigg) \Phi^{-1}(p)\sqrt{2\lambda_{t_{max}}RHIx } \geq 
  \lambda_{t} \Phi^{-1}(p)\sqrt{2\lambda_{t}RHIx } \Bigg( = n_P^{req}(t) \Bigg), \forall t.\\  
  \end{split}.
\end{equation}

Therefore, the minimum required fleet size $m^*$ equals the required fleet size in $t_{max}$ according to Equation \ref{eq:mopt}, i.e., $\arg\max_t \left( m^{req} (t) \right) = t_{max}$

When the demand rate is the lowest in $t_{min}$, the highest number of idle vehicles are parked, so the confidence level $p$ (the probability that there is at least one idle vehicle at the station nearest to a demand) can be set to 1 for further efficiency. On the other hand, during the peak hour $t_{max}$, the confidence level q can be reasonably assumed to be 1 in the time window $t_{max}$. Based on this, it is possible to simplify and rewrite the equations of $m^*$ by using the maximum passenger demand ($\lambda_{t_{max}}$) , as follows:

\begin{equation}
  \begin{split}
   m^*
     & = \lambda_{t_{max}} R \left( \frac{l_{t_{max}}}{v_{min}} \right) + \lambda_{t_{max}} R T^1_{A,t_{max}} (1+p+\alpha p -\alpha p^2) + \kappa\Phi^{-1}(p)\sqrt{2\lambda_{t_{max}} RHI}  \frac{1}{v_{min} T^1_{A,t_{max}}}
  \end{split},
  \label{eq:m_specialcase}
\end{equation}

Based on the operation depicted in Figure \ref{fig:scenario}, extra vehicles beyond the required fleet sizes for states $A$, $S$, and $C$ are not needed. In other words:

\begin{equation}
  \begin{split}
  & n_A(t) = n_A^{req}(t), n_S(t) = n_S^{req}(t), n_C(t) = n_C^{req}(t), \forall t \\
  \end{split},
\end{equation}

On the other hand, as shown in Figure \ref{fig:numveh}, the number of vehicles parked at parking stations in $t$, $n_P (t)$, is not always the same as $n_P^{req} (t)$, but can be found as Equation \ref{eq:npreq}.

\begin{equation}
  \begin{split}
  n_P(t)=m^* - \left( n_A(t) + n_S(t) + n_C(t)  \right)
  \end{split},
  \label{eq:npreq}
\end{equation}

The number of vehicles not parked in stations, $n_A(t) + n_S(t) + n_C(t)$, is the lowest in $t_{min}$, so the number of parked vehicles is the highest in $t_{min}$. The minimum required number of parking spaces $y^*R$ is the summation of the daily maximum number of required parking spaces, i.e., $n_P(t_{min})$, and additional buffer spaces to guarantee that each parking station is not full by confidence level $q$, $\Phi^{-1}(q)\sqrt{2\lambda_{t_{min}} RHIx}$. As a result, the optimal density of parking spaces can be found as shown in Equation \ref{eq:y_specialcase}

\begin{equation}
  \begin{split}
  y^* = & \frac{m^* - \left( n_A(t_{min}) + n_S(t_{min}) + n_C(t_{min}) \right) + \Phi^{-1}(q)\sqrt{2\lambda_{t_{min}} RHIx}}{R} \\ 
= & 
 \left(  \frac{\lambda_{t_{max}} l_{t_{max}}}{v_{min}} - \frac{\lambda_{t_{min}} l_{t_{min}}}{v_{max}}  \right) \\
&+ 
\left(
(1+p+\alpha p - \alpha p^2) \lambda_{t_{max}}  - (1+q+\alpha q - \alpha q^2) \lambda_{t_{min}} \cdot \frac{v_{min}}{v_{max}} 
\right) T^1_{A,t_{max}} \\
&+ \frac{\kappa}{v_{min}}\left(  \Phi^{-1} (p) \sqrt{\frac{2 \lambda_{t_{max}} H I}{R}}  + \frac{v_{min}}{v_{max}} \Phi^{-1} (q) \sqrt{\frac{2 \lambda_{t_{min}} H I}{R}} \right) \left( \frac{1}{T^1_{A,t_{max}}} \right)
\\
  \end{split}.
  \label{eq:y_specialcase}
\end{equation}

\subsection{Solution of Optimal Parking Planning}\label{sec:parkop}

The objective function in Equation \ref{eq:ob_con} is reformulated in terms of $T^1_{A,t_{max}}$ in Equation \ref{eq:ob_T_A_reform}, which is a function of the remaining sole decision variable $x$ (see $T^1_{A,t} = \frac{\kappa}{v_t \sqrt{x}}$ from Equation \ref{eq:xTa}). There are four terms in Equation \ref{eq:ob_T_A_reform}: the constant term, $\frac{1}{(T^1_{A,t_{max}})^2}$, $\frac{1}{T^1_{A,t_{max}}}$, and $T^1_{A,t_{max}}$.

\begin{equation}\label{eq:ob_T_A_reform}
\begin{split}
 & \min_{T_A}{  \left( 
  \begin{split}
&   \left( \left( C_y + C_m \right) \left( \frac{\lambda_{t_{max}} l_{t_{max}}}{v_{min}} \right) R - C_y\left( \frac{\lambda_{t_{min}} l_{t_{min}}}{v_{max}} \right) R \right)\\
  &+\frac{C_x\kappa^2 R}{(v_{min})^2} \frac{1}{(T^1_{A,t_{max}})^2}+\\
%
  & + \left( 
  \frac{\kappa(C_y+C_m)}{v_{min}} \Phi^{-1} (p) \sqrt{2 \lambda_{t_{max}} HIR } + \frac{\kappa C_y}{v_{max}} \Phi^{-1} (q) \sqrt{2 \lambda_{t_{min}} H I R }
  \right)  \frac{1}{T^1_{A,t_{max}}}\\
%
& + \left(
(C_y+C_m) \lambda_{t_{max}} R (1+p+\alpha p - \alpha p^2)
- C_y \lambda_{t_{min}} R (1+q+ \alpha q - \alpha q^2) \frac{v_{min}}{v_{max}}
\right) T^1_{A,t_{max}}
\\
  \end{split}
  \right)}  \\ 
   & \begin{split}
     s.t. T_{A,t_{max}} = T^1_{A,t_{max}} (1+p+\alpha p - \alpha p^2) \leq T_0, 
    \end{split} \\
\end{split}
\end{equation}

For simplicity of derivation, each coefficient in Equation \ref{eq:ob_T_A_reform} are parameterized as $P_0$, $P_{-2}$, $P_{-1}$, and $P_1$ as shown in Equation \ref{eq:ob_T_A_reform_param}.

\begin{equation}\label{eq:ob_T_A_reform_param}
  \begin{split}
  \min_{T^1_{A,t_{max}}} Cost \left( T^1_{A,t_{max}} \right) = \min_{T^1_{A,t_{max}}} \left({P_0 + P_{-2} \frac{1}{(T^1_{A,t_{max}})^2} + P_{-1} \frac{1}{T^1_{A,t_{max}}} + P_{1} T^1_{A,t_{max}}  }\right)
  \end{split},
\end{equation}

Since $\lambda_{max} > \lambda_{min}$ and all parameters are positive by definition, $P_0,P_{-2},P_{-1},P_{1} > 0$. Therefore, Equation \ref{eq:ob_T_A_reform_param} is a convex curve in the first quadrant, and has a local minimum point in the first quadrant. Consequently, we take the derivative of Equation \ref{eq:ob_T_A_reform_param} to find the unconstrained optimal point.

\begin{equation}\label{eq:ob_T_A_derivative}
  \begin{split}
  & \frac{d Cost \left( T^1_{A,t_{max}}  \right) }{d T^1_{A,t_{max}}} = -2P_{-2}(T^1_{A,t_{max}})^{-3} -P_{-1}(T^1_{A,t_{max}})^{-2}+P_1 = 0 \\
  & -2P_{-2} -P_{-1} \left(T^1_{A,t_{max}}\right)+P_1\left(T^1_{A,t_{max}}\right)^3 = 0 \\
  & \left( T^1_{A,t_{max}} \right)^3 - \frac{P_{-1}}{P_1} T^1_{A,t_{max}} - \frac{2P_{-2}}{P_1} = 0\\ 
  \end{split},
\end{equation}

For simplicity in derivation, let $A = \left( -\frac{P_{-1}}{P_{1}} \right)$ and $B = \left( -\frac{2P_{-2}}{P_{1}} \right)$. 
With realistic ranges of parameters, the discriminant ($\Delta$) of the cubic equation is positive, and there are three distinct real roots as shown in Equation \ref{eq:discriminant}:

\begin{equation}\label{eq:discriminant}
  \begin{split}
  \Delta & = -\left( 4A^3+27B^2 \right) > 0 \\
  \end{split},
\end{equation}

When there are three real roots in cubic equation, François Viète (1540-1603) derived the trigonometric solution. The three real roots ($t_k$) can be calculated as shown in Equation \ref{eq:viete}:

\begin{equation}\label{eq:viete}
  \begin{split}
  & t^3 + pt + q = 0\\
  & t_k = 2 \sqrt{-\frac{p}{3}} 
  \cos{\left(
  \frac{1}{3} \arccos{ \left(
  \frac{3q}{2p} \sqrt{-\frac{3}{p}}
  \right)} - k \frac{2 \pi}{3}
  \right)} \quad \mathrm{for} \quad k=0,1,2 \\
  \end{split}.
\end{equation}

The only positive solution is when $k=0$. As a result, 

\begin{equation}\label{eq:viete_sol}
  \begin{split}
  T_{A,t_{max}}^{1,u}
  &= 2 \sqrt{-\frac{A}{3}} 
  \cos{\left(
  \frac{1}{3} \arccos{ \left(
  \frac{3B}{2A} \sqrt{\frac{-3}{A}}
  \right)}
  \right)} \\
  &= 2 \sqrt{\frac{P_{-1}}{3P_{1}}} 
  \cos{\left(
  \frac{1}{3} \arccos{ \left(
  \frac{6P_{-2}}{2P_{-1}} \sqrt{\frac{3P_{1}}{P_{-1}}}
  \right)}
  \right)} \\  
  \end{split}.
\end{equation}




Equation \ref{eq:viete_sol} is the unconstrained minimum of Equation \ref{eq:ob_T_A_reform}. Thus, if $T_0 > (p+\alpha p - \alpha p^2) T_{A,t_{max}}^{1,u}$, the constraint is not binding since $T_{A,t_{max}} \geq T_{A,t}$, so that the optimal point $T_{A,t_{max}}^{1,*}=T_{A,t_{max}}^{1,u}$ and $Cost^* = Cost\left(T_{A,t_{max}}^{1,u} \right)$. On the other hand, if $T_0 \leq (p+\alpha p - \alpha p^2) T_{A,t_{max}}^{1,u}$, the constraint is binding, so that the optimal point $T_{A,t_{max}}^{1,*} = \frac{T_0}{(p+\alpha p - \alpha p^2)}$ and $Cost^* = Cost \left(\frac{T_0}{(p+\alpha p - \alpha p^2)} \right)$:

\begin{equation}
  \begin{split}
  T_{A,t_{max}}^{1,*} = \begin{cases}
  &T_{A,t_{max}}^{1,u}, \quad \textrm{if} \quad T_0 > (1+p+\alpha p - \alpha p^2 )T_{A,t_{max}}^{1,u} \\
  &\\
  &\frac{T_0}{(1+p+\alpha p - \alpha p^2)}, \quad \textrm{otherwise}\\
  \end{cases}\\   
  \end{split}.
\end{equation}

Then, based on the optimal decision variable ($T_{A,t_{max}}^{1,*}$), we can calculate three parking operational variables ($x$, $m$, and $y$) by using Equation \ref{eq:xTa}, Equation \ref{eq:m_specialcase}, and Equation \ref{eq:y_specialcase}, respectively. Since we have the analytical form of the solution, we can calculate parking operational variables in one shot.

\section{Case Study for Single-Zone Analytical Parking Planning Model}\label{sec:case_study_sappm}

In this section, we demonstrate S-APPM through a case study in Seoul, South Korea. Specifically, we will discuss changes in the density of parking stations and parking spaces as well as fleet size required to serve passenger demand when system is optimized. The demand distribution in Seoul is not spatially uniform, but the results, assuming uniformity, can give an upper bound of system efficiency, which is useful at the beginning of high-level planning. The next model, T-APPM, which will be elaborated in Section 5, can be easily extended to a general multi-zonal framework. To realistically account for the spatial demand heterogeneity with zonal-specific parking planning, advanced models can be used instead of S-APPM to provide the upper bound of the reality.


\begin{table}[!ht]
	\caption{Model parameters}
	\begin{center}
	\begin{tabular}{l|l|l}
	     Variable &  Units & Value  \\\hline\hline
	     $R$ & $[\mathrm{km^2}]$ & 605.24\\\hline
	     $l$  & $[\mathrm{km}]$ & 16.4 \\\hline
	     $v_{min}$  & $[\mathrm{km/hr}]$ & 18.0 \\\hline
	     $v_{max}$ & $[\mathrm{km/hr}]$ & 40.0  \\\hline
	     $H$  & $[\mathrm{hr}]$ & 2  \\\hline
	     $p$ & - & 0.95 \\\hline
	     $q$ & - & 0.95  \\\hline
	     $\alpha$  & -  & 2 \\\hline
	     $I$  & - & 1 \\\hline
	     $\kappa$  & - & 0.5 \\\hline
	     $T_0$  & $[\mathrm{min}]$ & 1   \\\hline
	     \hline
	\end{tabular}
	\end{center}
	\label{tab:s-appm_parameters}
\end{table}

Table \ref{tab:s-appm_parameters} shows the parameters used in this case study. The area of Seoul is approximately $605.2 \mathrm{km^2}$. According to the Seoul Travel Survey, the average trip length is $16.4 \mathrm{km}$, and the minimum and maximum speeds are $18.0 \mathrm{km/hr}$ and $40.0 \mathrm{km/hr}$, respectively. We consider discrete time windows, each of which has a length of two hours. Both confidence levels, to guarantee that there is at least one vehicle at each parking station ($p$) and that there is at least one parking space left at each parking station ($q$), are set at 0.95. The incremental ratio of travel time to the second nearest parking station ($\alpha$) is 2. The mean-to-variance ratio (I) is set to 1 assuming that occurrences of O-D events follow a Poisson distribution. Finally, $\kappa$ is set to 0.5 referenced from \cite{daganzo2019public}.

\begin{table}[!h]
	\caption{Corresponding decision variables in the current transportation system in Seoul}
	\begin{center}
	\begin{tabular}{l|l|l}
	     Variable & Unit & Value \\\hline\hline
	     $x_{Seoul}$& $[\mathrm{stations/km^2}]$ & 524.06  \\\hline
	     $y_{Seoul}$& $[\mathrm{spaces/km^2}]$  & 7,150.24 \\\hline
	     $z_{Seoul}$& $[\mathrm{spaces/stations}]$  & 13.64 \\\hline
	     $m_{Seoul}$& $[\mathrm{veh}]$ & 2,703,429  \\\hline
	     \hline
	\end{tabular}
	\end{center}
	\label{tab:seoul_var}
\end{table}


It is worth investigating actual numbers for each decision variable in Seoul. According to the statistics on parking spaces in Seoul, there are 317,181 parking stations and 4,327,614 parking spaces, including public parking stations, private parking stations, and residential parking stations. Therefore, as shown in Table \ref{tab:seoul_var}, the density of parking stations in Seoul is 524.09 $\mathrm{stations/km^2}$, and the density of parking spaces in Seoul is 7150.72 $\mathrm{spaces/km^2}$. There are approximately 13.64 parking spaces at each parking station. Moreover, there are 3,157,361 registered passenger vehicles in Seoul, which means that there are 0.625 parking spaces for each vehicle.

\begin{table}[!ht]
	\caption{Hourly average passenger demand in Seoul. The values are in $[veh/km^2/hr]$}
	\begin{center}
	\begin{tabular}{l|p{3cm}|p{3cm}}
    Time Window & Passenger Demand by Personal Vehicle & Passenger Demand by All Mode
    \\\hline\hline 
    Total & 285.11 & 1720.03 \\\hline
    AM peak (7-9 AM) &  765.04 & 4518.16\\\hline
    PM peak (6-8 PM) &  836.94 & 4042.69 \\\hline
    Off peak  &  181.93 & 1207.95 \\\hline
	     \hline
	\end{tabular}
	\end{center}
	\label{tab:seoul_demand}
\end{table}

Table \ref{tab:seoul_demand} shows the average hourly unit passenger demand (in [$veh/km^2/hr$]) in Seoul according to the National Household Travel Survey (O-D Flow Survey) in Korea. This survey offers the average hourly flow from one district to another, classified by mode of transportation and purpose of trip. The mode of transportation in this survey includes passenger vehicles, buses, subways, high-speed rail, walking, and bicycles; we extracted passenger demand by personal vehicle and passenger demand by all modes in the different time windows. Table \ref{tab:seoul_demand} shows the corresponding values in each time window. The values in the total row represent the average passenger demand during any time-of-day. The values in the \textit{AM peak} row represent the average passenger demand during the morning peak (7-9 AM); the values in \textit{PM peak} row represent the average passenger demand during the afternoon peak (6-8 PM). In this study, we assume that passenger demand in time windows other than AM and PM peaks is equal to off-peak passenger demand. As a result, the values in the \textit{Off-peak} row were calculated based on the values in the Total, AM peak, and PM peak rows.

\begin{table}[!ht]
	\caption{Range of cost variables used in the sensitivity analysis.}
	\begin{center}
	\begin{tabular}{l|l|l}
	     Cost & Value & Reference \\\hline\hline 
	     $C_m$ &  $[30,31,\cdots,\textbf{35.616},\cdots,\textbf{183.36},\cdots,200]$ & \cite{estrada2021operational} \\\hline
	     $C_x$ &  $[0.1,0.2,\cdots,4.9,5.0]$ & - \\\hline
 	     $C_y$ &  $[0.1,0,2,\cdots,\textbf{4.73},\cdots,20]$ & Korean Ministry of Land, Infrastructure, and Transport \\\hline
	     \hline
	\end{tabular}
	\end{center}
	\label{tab:cost}
\end{table}

In this case study, we analyze the results of our proposed model depending on different values for three operational costs: $C_m$, $C_x$, and $C_y$. The values and the references used for the sensitivity analysis are shown in Table \ref{tab:cost}. 
The daily operation cost of each SAV fleet ($C_m$) is referenced from \cite{estrada2021operational}, which analyzed the operational cost of on-demand bus and taxi service, presenting the operational cost for different types of powertrain (diesel and electric) and different sizes of vehicle (standard bus, mini-bus, and passenger car). In this study, we assume that each vehicle serves one passenger demand for one operation and SAVs use an electric powertrain; we use the \textit{electric passenger car} as reference vehicle for our study. In \cite{estrada2021operational} there are two reference values with (35.616) and without (183.36) driver expenses; as a result, we set the range of $C_m$ from 30 to 200 $\$/veh/day$.
%
%
%
The daily operation cost of the parking station ($C_x$) includes costs, such as amortized installation cost of the parking management system in the parking station. The daily operation cost of parking spaces ($C_y$) includes the amortized cost of purchasing the land used for the parking spaces and the construction cost for parking spaces. There are not many references to support our reasoning for choosing an appropriate value for $C_x$, but $C_x$ must be smaller than or similar to $C_y$ because the installation cost of the parking management system is not expensive compared to the overall land price for the parking spaces. 
We referenced one value for $C_y$ from the \textit{declared land value} announced by the Korean Ministry of Land, Infrastructure, and Transport. The average land price in Seoul is 2490.35 $\$/m^2$, and the average area of each parking space is 3.3 $m^2$. 
We assume that the land price is equivalent to 5-year rental cost with 2\% annual interest rate. 
As a result, the amortized daily cost for each parking space is $\left(  \frac{2490.35\cdot3.3\cdot\frac{0.02}{365}}{1-\left(1+\frac{0.02}{365}\right)^{-365\cdot5}}  \right)$ $=4.73\$/spaces/day$ considering only the land price. We set the range of $C_y$ from 0.1 to 20 $\$/spaces/day$ for the sensitivity analysis
Finally, since $C_x$ should be similar to or less than $C_y$, we set the range of $C_x$ from 0.0 to 5.0 $\$/stations/day$.
%

\begin{figure}[!t]
\centering
  \includegraphics[width=\textwidth]{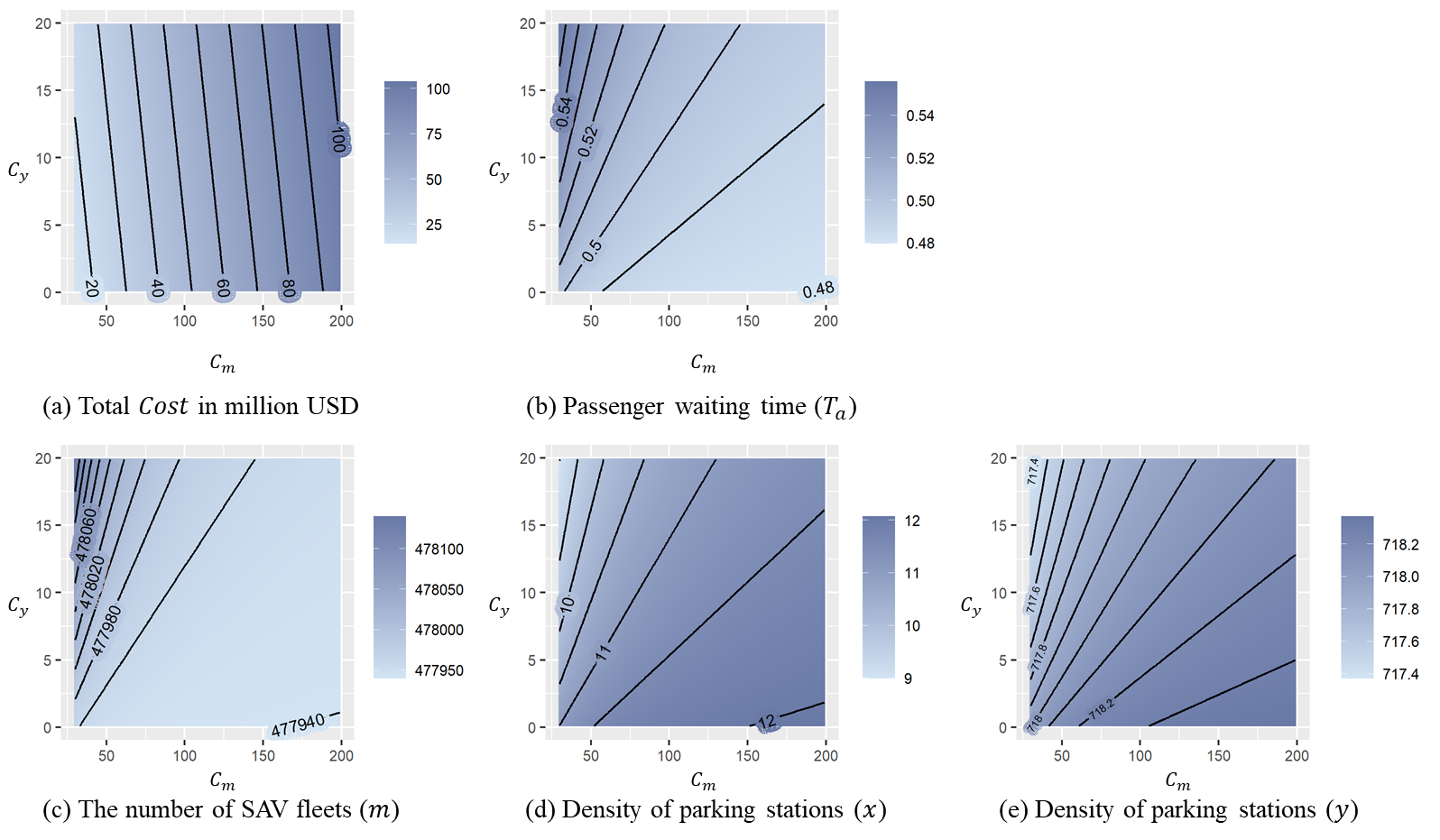}%
  \caption{Result of the sensitivity analysis between $C_m$ and $C_y$, when $C_x=2\$/stations/day$}
  \label{fig:sen_cmcy}
\end{figure}

\begin{figure}[!t]
\centering
  \includegraphics[width=\textwidth]{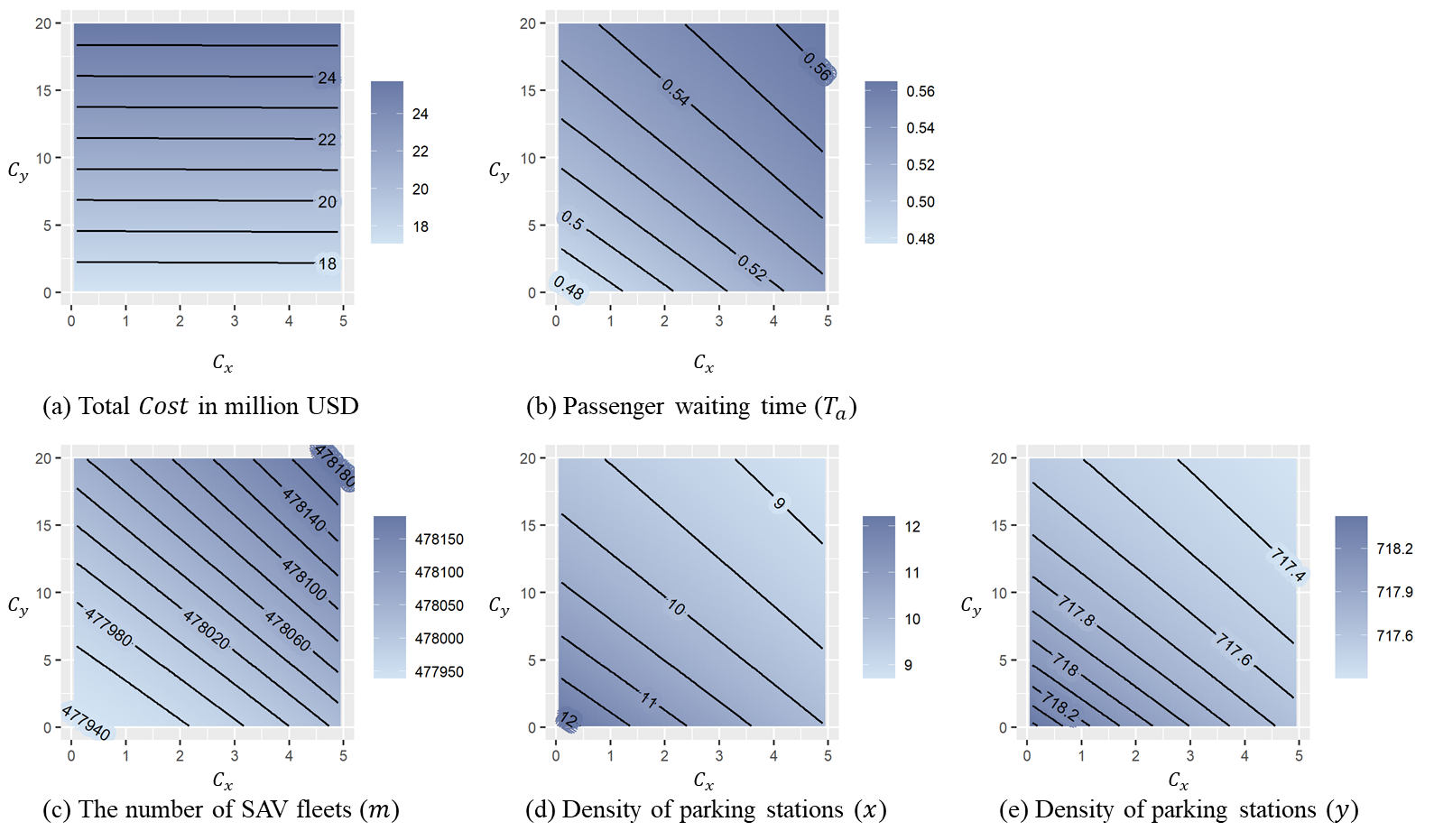}%
  \caption{Results of sensitivity analysis between $C_x$ and $C_y$, when $C_m=35.616\$/spaces/day$}
  \label{fig:sen_cxcy}
\end{figure}

Figure \ref{fig:sen_cmcy} and Figure \ref{fig:sen_cxcy} show the results of sensitivity analysis. The passenger demand by personal vehicle in Table \ref{tab:seoul_demand} is used for the sensitivity analysis. Figure \ref{fig:sen_cmcy} shows the result with different $C_m$ and $C_y$ when $C_x$ is fixed to $2 \$/stations/day$. Figure \ref{fig:sen_cxcy} shows the result with different $C_x$ and $C_y$ when $C_m$ is fixed to $35.616 \$/veh/day$.
There are six results of different variables in each figure: Total cost in million USD ($Cost$), Passenger waiting time ($T_A=T_{A,t_{max}}^1 (p+\alpha p - \alpha p^2)$) in minutes, the number of SAV fleets ($m$), density of parking stations ($x$), density of parking spaces ($y$), and the average number of parking spaces in each parking station ($z$).

Figure \ref{fig:sen_cmcy} (a) shows that the total cost significantly drops as $C_m$ decreases. 
This result shows that replacing human-driven taxis with SAVs will significantly improve cost-efficiency. 
In Figure \ref{fig:sen_cmcy} (b), when $C_m$ is fixed, there is an increasing tendency in $T_A$ as $C_y$ increases. On the other hand, there is a decreasing tendency in $T_A$ as $C_m$ increases when $C_y$ is fixed. 
Within the realistic range of each cost value, the unconstrained minimum of Equation \ref{eq:ob_T_A_reform_param} ($T_{A,t_{max}}^{1,u}$) is not bounded by the constraint, $T_0\leq (p+\alpha p -\alpha p^2)T_{A,t_{max}}^{1,u}$, so the optimal value for $T_{A,t_{max}}^1$ is equal to $T_{A,t_{max}}^{1,u}$.
As a result, three operational variables can be calculated from the derived Equations in Section \ref{section:2.1.2}.

The number of SAV fleets ($m$) is calculated based on Equation \ref{eq:m_specialcase}; the results are shown in Figure \ref{fig:sen_cmcy} (c). Similar to $T_A$, there is an increasing tendency in $m$ as $C_y$ increases and there is a decreasing tendency in $m$ as $C_m$ increases.
Furthermore, the density of parking spaces ($x$) is proportional to the square of reciprocal of $T_A$ as shown in Equation \ref{eq:xTa}. As a result, there is a decreasing tendency in $x$ as $C_y$ increases and there is an increasing tendency in $x$ as $C_m$ increases as shown in Figure \ref{fig:sen_cmcy} (d). 
Within the range of $T_A$ shown in Figure \ref{fig:sen_cmcy} (a), the $T_A$ term in Equation \ref{eq:y_specialcase} is relatively smaller than $\frac{1}{T_A}$ term. As a result, $y$ has a tendency opposite to $T_A$ as shown in Figure \ref{fig:sen_cmcy} (e). 

Figure \ref{fig:sen_cxcy} (a) shows the change in total cost with different values of $C_x$ and $C_y$; the overall cost decreases as $C_y$ decreases.
In Figure \ref{fig:sen_cxcy} (b), when $C_x$ is fixed, there is an increasing tendency in $T_A$ as $C_y$ increases. Similarly, as $C_x$ increases, there is an increasing tendency in $T_A$. When the cost of parking facilities is low, it is relatively cost-efficient to install more parking stations and parking spaces. The passenger waiting time consequently decreases as the number of parking stations increases. 
Similar to the previous results in Figure \ref{fig:sen_cmcy}, $m$ follows a similar tendency with $T_A$, while $x$ and $y$ follow a tendency opposite to that of $T_A$.

%

Table \ref{tab:result_summary} shows a summary of the results for different levels of demand. The upper part of the table shows the results when SAVs serve current passenger demand for personal vehicles; the lower part shows the results when SAVs serve passenger demand for all modes, including private vehicles and public transportation. To derive these results, we used $C_m = 35.616$ to show that all vehicles are autonomous vehicles without human drivers, and $C_y = 4.73$ to represent the land price in Seoul. $C_x = 2$ was arbitrarily chosen.

\begin{table}[b]
	\caption{Summary of results of Case Study for S-APPM}
	\begin{center}
	\begin{tabular}{l|ll|l|ll}
	     Demand & Variable&  & Current & Optimal & \\ \hline\hline
	     \multirow{6}{*}{Personal Vehicle} 
	     &$x$ & [$\mathrm{stations/km^2}$] & 524.09 & 11.66 &(\color{red} $-97.75\%$ ) \\\cline{2-6}
	     &$y$& [$\mathrm{spaces/km^2}$] & 7,150.72 & 718.22& (\color{red} $-89.96\%$) \\\cline{2-6}
	     &$z$ & [$\mathrm{spaces/station}$] &13.64 & 61.59 & (\color{blue} $+351.54\%$) \\\cline{2-6}
	     &$m$ & [$\mathrm{veh}$] & 2,703,429 & 477,944.71 &(\color{red} $-82.32\%$) \\\cline{2-6}
 	     &$yR/m$ & [$\mathrm{spaces/veh}$] & 1.601 & 0.9095 &(\color{red} $-43.19\%$) \\\cline{2-6}
	     \hline\hline
	     \multirow{6}{*}{All Mode} 
     &$x$ & [$\mathrm{stations/km^2}$]  & 524.09 & 27.51 &(\color{red} $-94.75\%$ ) \\\cline{2-6}
     &$y$ & [$\mathrm{spaces/km^2}$] & 7,150.72 & 3728.66  & (\color{red} $-47.86\%$) \\\cline{2-6}
     &$z$ & [$\mathrm{spaces/station}$] & 13.64 & 135.52 & (\color{blue} $+893.55\%$) \\\cline{2-6}
     &$m$ & [$\mathrm{veh}$] & 2,703,429 & 2,549,647.66 &(\color{red} $-5.69\%$) \\\cline{2-6}
      &$yR/m$ & [$\mathrm{spaces/veh}$] & 1.601 & 0.8851 &(\color{red} $-44.72\%$) \\\cline{2-6}
	     \hline\hline
	\end{tabular}
	\end{center}
	\label{tab:result_summary}
\end{table}

The values in the column named ``Current'' show the values corresponding to the current transportation system in Seoul; the values in the column marked ``Optimal'' show results of S-APPM when the SAV system replaces the current transportation system. The results show that it is possible to significantly decrease the number of parking stations, parking spaces, and vehicles by introducing the SAV system. The density of parking stations decreases 97.75\% and the density of parking spaces for personal vehicle demand decreases 89.96\%. It is notable that the average number of parking spaces at each parking station ($z$) increases from 13.64 to 61.59. The current system in Seoul has fewer parking spaces in one parking station and the parking stations are dense. However, the optimal solution for the SAV system suggests that sparse parking stations with more parking spaces at each station will be more cost-efficient. The fleet size ($m$) also significantly decreases. Approximately 5.66 personal vehicles can be replaced with one SAV. Finally, the number of parking spaces for each vehicle decreases from 1.601 to 0.9095. There is less than one parking space for one vehicle in the optimal solution. This means that at least 9.05\% of vehicles are out on the road serving passengers. This is because of the demand setting. The passenger demand at off-peak is assumed to be the same at any time window other than those at AM-peak and PM-peak. With better data on passenger demand, this result can realistically be improved.

When passenger demand increases to the demand by all mode, as shown in Table \ref{tab:result_summary}, all three operational variables increase corresponding to the increase in demand. Although the density of parking stations increased, the density of parking spaces increased more significantly. As a result, the average number of parking spaces at each parking station ($z$) increased too. This again shows that sparse parking stations with more parking spaces at each station will be more cost-efficient.

\clearpage

\clearpage

\clearpage  
\section{Two-Zone Analytical Parking Planning Model}\label{sec:tappm}

\begin{figure}[!t]
  \centering
  \includegraphics[width=0.7\textwidth]{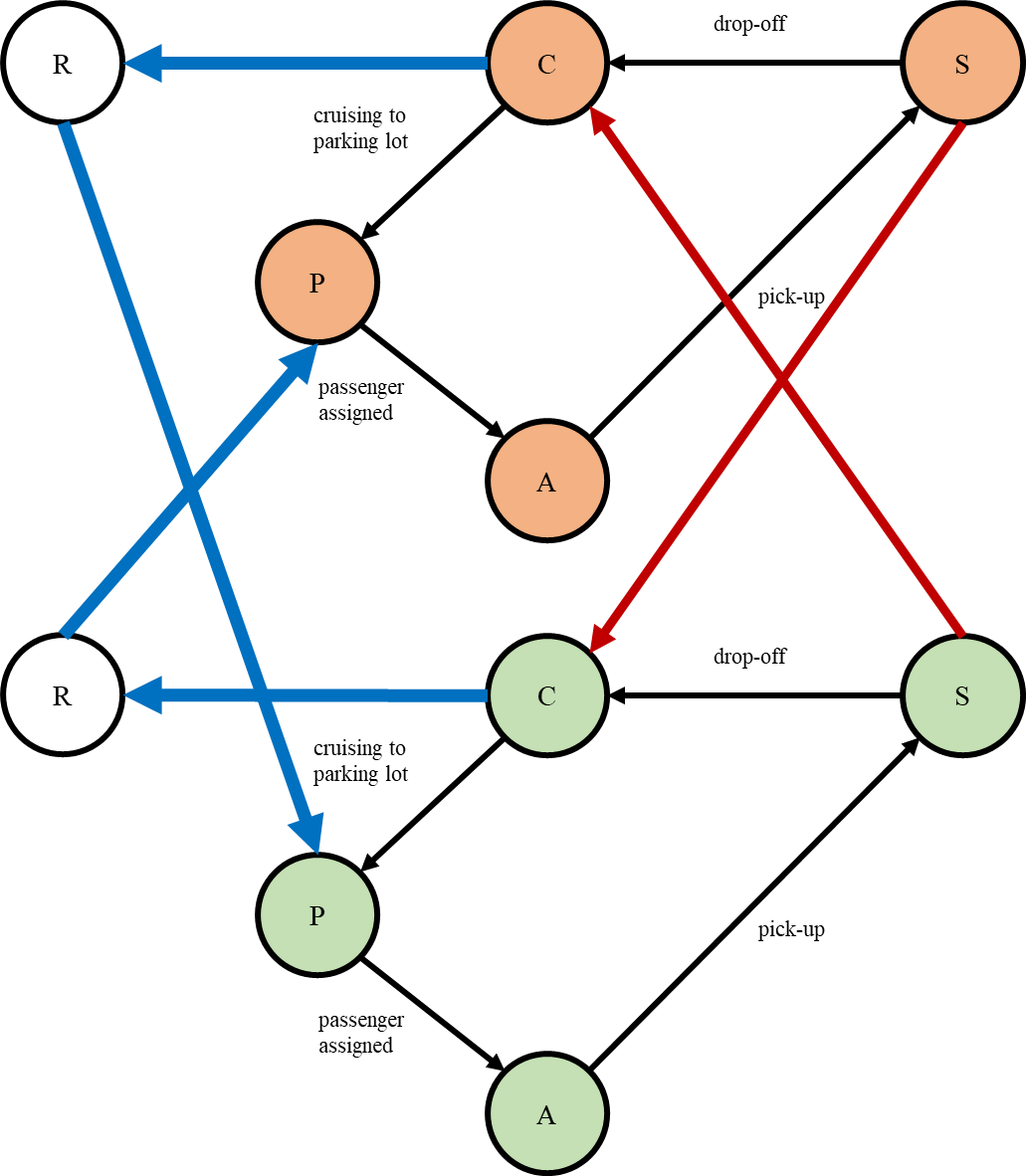}
  \caption{Workload representation of Two-zone Analytical Parking Planning Model (T-APPM)}\label{fig:twozonemodel}
\end{figure}


In this section, we model the \textit{Two-zone Analytical Parking Planning Model (T-APPM)} by extending S-APPM, as shown in Figure \ref{fig:twozonemodel}. Since T-APPM considers the target region as two distinguishable zones, the model must consider inter-zonal movements. First, the origin and destination of passenger trips can be located in different zones. The red lines in Figure \ref{fig:twozonemodel} indicate passenger trips from one zone to another. Second, it is necessary to consider relocation of SAVs. When the passenger demand is higher in one direction than the opposite, the number of SAV fleets (both parked and running) in one zone will continuously increase. For example, in the morning peak, passenger demand from suburb to city center will be much higher than passenger demand from city center to suburb. When this demand pattern continues, there will be more vehicles in the city center and fewer vehicles in the suburb, causing a lack of parking spaces in the city center and a decrease in level-of-service in the suburb by increasing the passenger waiting time. As a result, a proper relocation strategy for SAVs should be considered. The blue lines in Figure \ref{fig:twozonemodel} describes the flow of vehicles that are relocated to the other zone after finishing passenger trips.

\begin{figure}[!t]
\centering
\begin{subfigure}{.5\textwidth}
  \centering
  \includegraphics[width=0.7\textwidth]{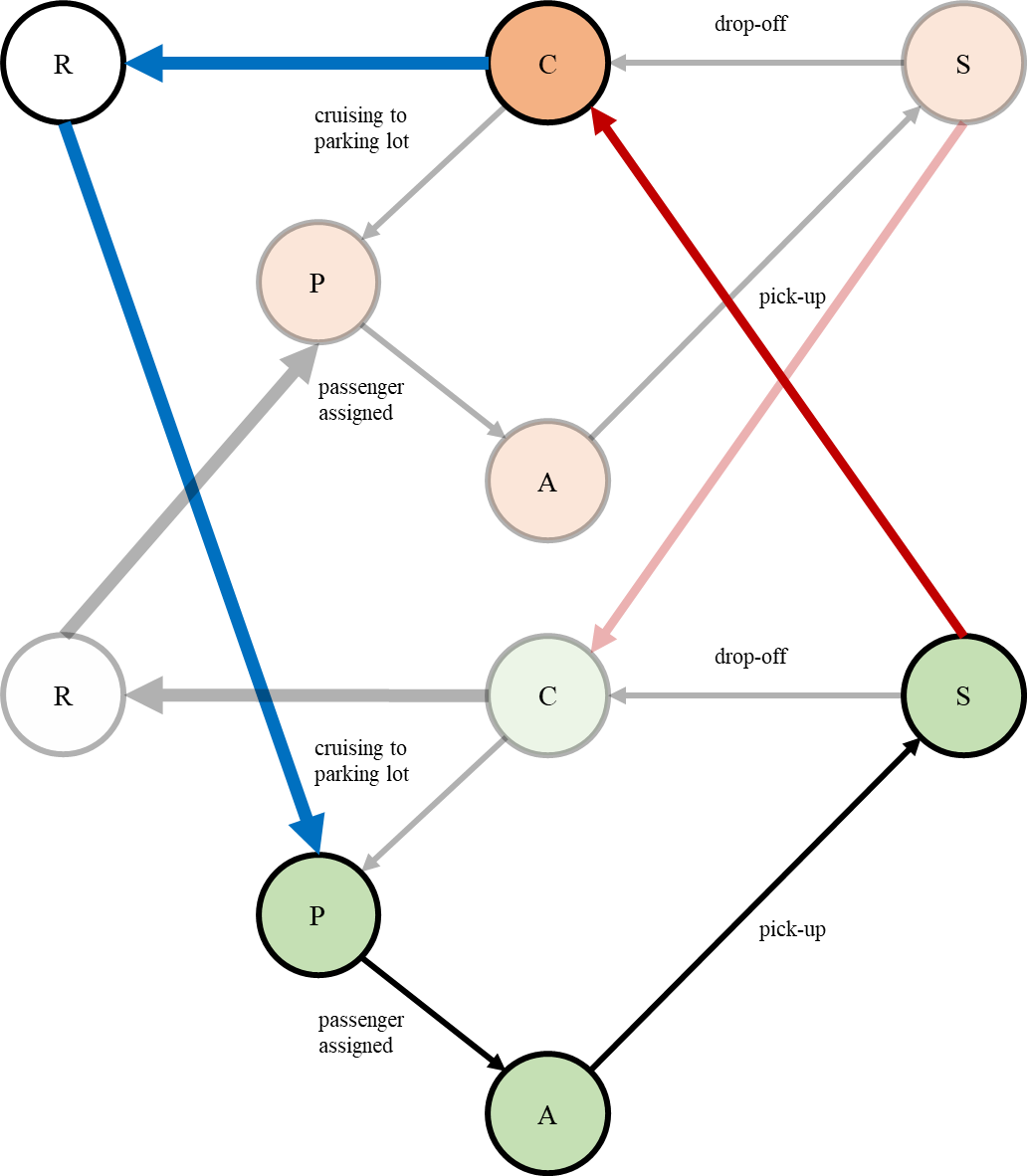}%
  \caption{}\label{fig:twozonemodel_highlight}%
\end{subfigure}%
\hfill%
\begin{subfigure}{.5\textwidth}
\centering
  \includegraphics[width=0.7\textwidth]{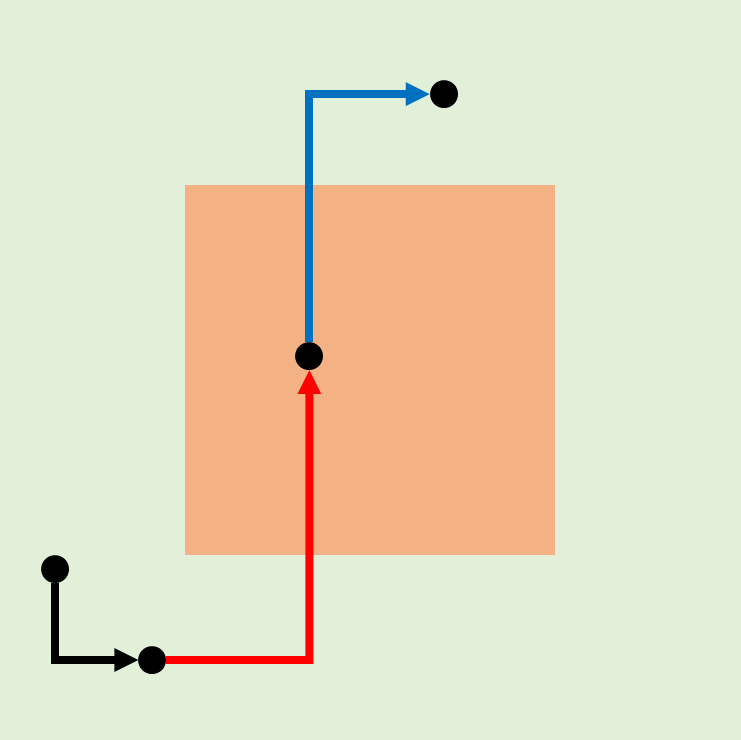}%
  \caption{}\label{fig:twozonescenario}%
\end{subfigure}%
\caption{Example case of T-APPM. Black line represents the vehicle traveling from the parking station to passenger’s origin, red line represents the passenger trip, and blue line represents the relocation of the vehicle. }
\label{fig:twozone_problem}
\end{figure}


Figure \ref{fig:twozone_problem} shows an example of T-APPM. A passenger calls for a trip from a suburb (colored in green) to the city center (colored in orange). The black line in Figure \ref{fig:twozone_problem} (b) represents the movement of the assigned vehicle from the parking station to the passenger’s origin. This corresponds to the state transition from $P$ to $A$ and the state transition from $A$ to $A$ in S-APPM of the suburb zone. The red line in Figure \ref{fig:twozone_problem} (b) represents the passenger trip from the suburb to the city center, which corresponds to the state transition from $S$ in the suburb to $C$ in the city center. Then, this vehicle is relocated to the suburb, which is represented by the blue line in Figure \ref{fig:twozone_problem} (b). This corresponds to the state transition from $C$ in the city center to $R$ (relocation state), and the state transition from $R$ to $P$ in the suburb.

Consider a given target region divided into two separate zones, as shown in Figure \ref{fig:twozone_problem} (b). The inner zone (zone 1) colored in orange has a size of $R_1$; the outer zone (zone 2), colored in green has a size of $R_2$. In zone 1, the density of parking stations is denoted as $x_1$, and the density of parking spaces is denoted as $y_1$. Likewise, the density of parking stations in zone 2 is denoted as $x_2$ and the density of parking stations in zone 2 is denoted as $y_2$. The total number of SAV fleets is denoted as $M=m_1+m_2$, where $m_1$ and $m_2$ represents the number of SAV fleets in zone 1 and zone 2, respectively. Similar to the assumptions in S-APPM, when passenger demand is generated, an SAV in the nearest not-empty parking station is assigned to the passenger ($P\rightarrow A$). This SAV cruises to the origin of the passenger and picks up the passenger ($A \rightarrow S$). The SAV travels from the origin to the destination and drops off the passenger ($S \rightarrow C$). If the SAV is relocated to a different zone ($C \rightarrow R$), the vehicle moves to the nearest parking station in the relocated zone ($R \rightarrow P$). Otherwise, the SAV moves to a parking station in the same zone from the destination of the passenger ($C \rightarrow P$)


\subsection{Objective Function}
The objective function of T-APPM is to minimize the total operation cost in the target regions: i) parking station operation costs in both zones, ii) parking space operation costs in both zones, and iii) fleet operation costs. The parking station operation cost ($C_x$) and fleet operation cost ($C_m$) are not different across two zones. However, the parking space operation cost ($C_y$) can be different across two zones because it includes the land cost. As a result, we assume that two zones have different cost values for $C_y$, but they have the same cost values for $C_x$ and $C_m$. 
The objective function $J$ is to minimize overall daily average operation cost ($Cost$) formulated as a function of the planning variables ($x_1,x_2,y_1,y_2,M$) with respect to LOS constraint:

\begin{equation}\label{eq:ob_tappm}
  \begin{split}
  & J = \min_{x_1,x_2,y_1,y_2,M}  Cost (x_1,x_2,y_1,y_2,M) = \min_{x_1,x_2,y_1,y_2,M} {\Big(C_x (x_1R_1+x_2R_2) + C_{y,1} y_1R_1 + C_{y,2} y_2R_2 + C_m M \Big)} \\ 
  & \textrm{s.t.} T_{A,t} \leq T_0 \\
  \end{split},
\end{equation}

\noindent
where $C_{y,1}$ is the unit cost for each parking space in zone 1, and $C_{y,2}$ is the unit cost for each parking space in zone 2. $x_1$ and $x_2$ represent the density of parking stations in each zone, $y_1$ and $y_2$ represent the density of parking stations in each zone, and $M$ represents the total number of SAV fleets for the operation. $T_{A,t}$ is the average passenger waiting time in time window indexed by $t$, and $T_0$ is the threshold, the maximum allowed average passenger waiting time.

\subsection{SAV Operation Model with Parking and Relocation}

For a given time window ($t$), we define a passenger demand matrix as $\mathbf{\Lambda}^t$ as follows:

\begin{equation}
  \begin{split}
  \mathbf{\Lambda_t} = \begin{bmatrix}
    \lambda_{t,11} & \lambda_{t,12} \\
    \lambda_{t,21} & \lambda_{t,22} 
  \end{bmatrix}
  \end{split},
\end{equation}

\noindent
where $\lambda_{t,ij}$ refers to the unit passenger demand from zone $i$ to zone $j$ in the given time window $t$. It is assumed that the origin of passenger demand is uniformly distributed in zone $i$ and the destination of passenger demand is uniformly distributed in zone $j$.

The total number of SAV fleets can be derived by summing the numbers of vehicles at all states. Similar to the derivations in Section \ref{SAPPM}, we first calculate the ``required'' number of SAV fleets by summing the required number of SAV fleets in each state to ensure the level-of-services as follows:

\begin{equation}
  \begin{split}
  M^{req}(t) = & m^{req}_{1}(t) + m^{req}_{2}(t)\\
  =& \Big( n_{1,A}^{req}(t) + n_{1,S}^{req}(t) + 
  n_{1,C}^{req}(t) + 
  n_{1,P}^{req}(t)+ 
  n_{1,R}^{req}(t) \Big) \\ 
  &+ \Big( n_{2,A}^{req}(t) + n_{2,S}^{req}(t) + 
  n_{2,C}^{req}(t) + 
  n_{2,P}^{req}(t) + 
  n_{2,R}^{req}(t) \Big)
  \end{split},
\end{equation}

\noindent
where the first part($i$) of the subscript of $n_{i,X}$ refers to the zone, and the second part($X$) refers to the state of the vehicle. For example, $n_{2,P}$ refers to the number of SAV fleets in ``Parked'' state in Zone 2. Then, the number of SAV fleets ($M^*$) is the maximum of $M^{req}$

\begin{equation}
    \begin{split}
        M^* = \max_t (M^{req}(t))
    \end{split}
\end{equation}








The number of SAV in states $A$, $S$, $C$, and $P$ can be calculated based on the derivations from S-APPM. First, we can calculate the number of vehicles in state $A$ as shown in Equation  \ref{eq:TAPPM_nA}. Here, $T^1_{A,t}$ is assumed to be same in both zones.

\begin{equation}
    \label{eq:TAPPM_nA}
  \begin{split}
  & n_{1,A} ^{req}(t) = (\lambda_{t,11}+\lambda_{t,12}) R_1 T^1_{A,t} (p + \alpha p - \alpha p^2) \\
  & n_{2,A} ^{req}(t) = (\lambda_{t,22}+\lambda_{t,21}) R_2 T^1_{A,t} (p + \alpha p - \alpha p^2) 
  \end{split}.
\end{equation}

Second, the number of SAV in state $S$ can be calculated as shown in Equation \ref{eq:TAPPM_nS}.

\begin{equation}
\label{eq:TAPPM_nS}
  \begin{split}
  & n_{1,S} ^{req}(t) = \lambda_{t,11} R_1 \left( \frac{l_{t,11}}{v_{t,11}} \right) + \lambda_{t,12} R_1 \left( \frac{l_{t,12}}{v_{t,12}} \right) \\
  & n_{2,S} ^{req}(t) = \lambda_{t,22} R_2 \left( \frac{l_{t,22}}{v_{t,22}} \right) + \lambda_{t,21} R_2 \left( \frac{l_{t,21}}{v_{t,21}} \right)
  \end{split},
\end{equation}

\noindent
where $l_{t,ij}$ is the average trip length from zone $i$ to zone $j$ in time window $t$, and $v_{t,ij}$ is the average speed from zone $i$ to zone $j$ in time window $t$.

Third, the number of SAV fleets in state $C$ can be calculated as shown in Equation \ref{eq:TAPPM_nC}:
\begin{equation}
    \label{eq:TAPPM_nC}
  \begin{split}
  & n_{1,C} ^{req}(t) = (\lambda_{t,11}R_1+\lambda_{t,21}R_2) T^1_{C,t} (q+\alpha q - \alpha q^2)\\
  & n_{2,C} ^{req}(t) = (\lambda_{t,22}R_2+\lambda_{t,12})R_1 T^1_{C,t} (q+\alpha q - \alpha q^2)
  \end{split}.
\end{equation}
\noindent where $T^1_{C,t} = T^1_{A,t}$.

Next, the required number of SAV fleets in state $P$ can be calculated as shown in Equation \ref{eq:TAPPM_nPreq}. 

\begin{equation}
\label{eq:TAPPM_nPreq}
  \begin{split}
  & n_{1,P}^{req}  (t)  = \Phi^{-1} (p) \sqrt{2(\lambda_{t,11} + \lambda_{t,12}) R_1 H I x_1}\\
  & n_{2,P}^{req}  (t)  = \Phi^{-1} (p) \sqrt{2(\lambda_{t,22} + \lambda_{t,21}) R_2 H I x_2}
  \end{split},
\end{equation}

\begin{figure}[!t]
    \centering
    \begin{subfigure}{.5\textwidth}
      \centering
      \includegraphics[width=0.7\textwidth]{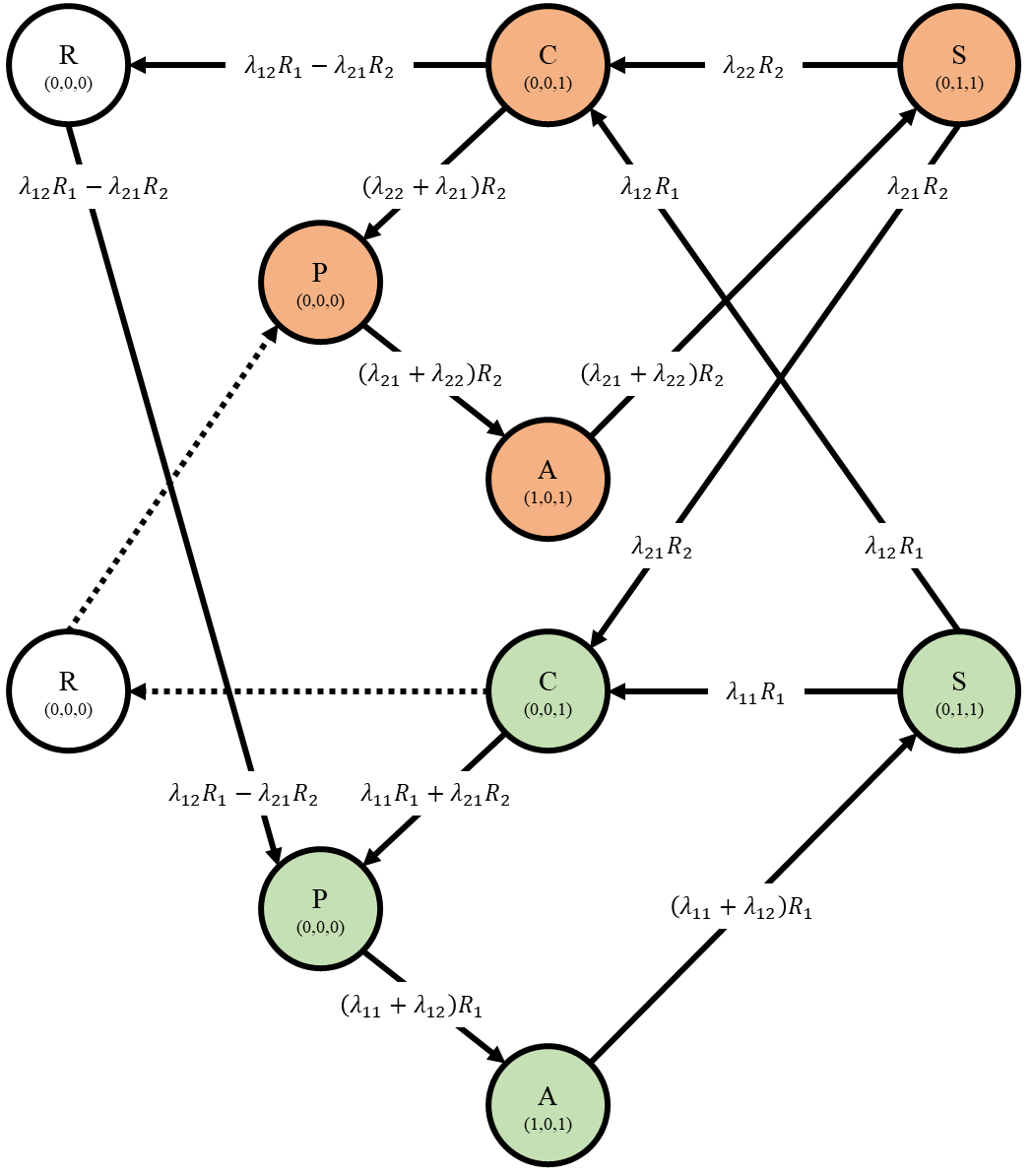}%
      \caption{$\lambda_{12}R_1 > \lambda_{21}R_2$}\label{fig:1221}%
    \end{subfigure}%
    \hfill%
    \begin{subfigure}{.5\textwidth}
    \centering
      \includegraphics[width=0.7\textwidth]{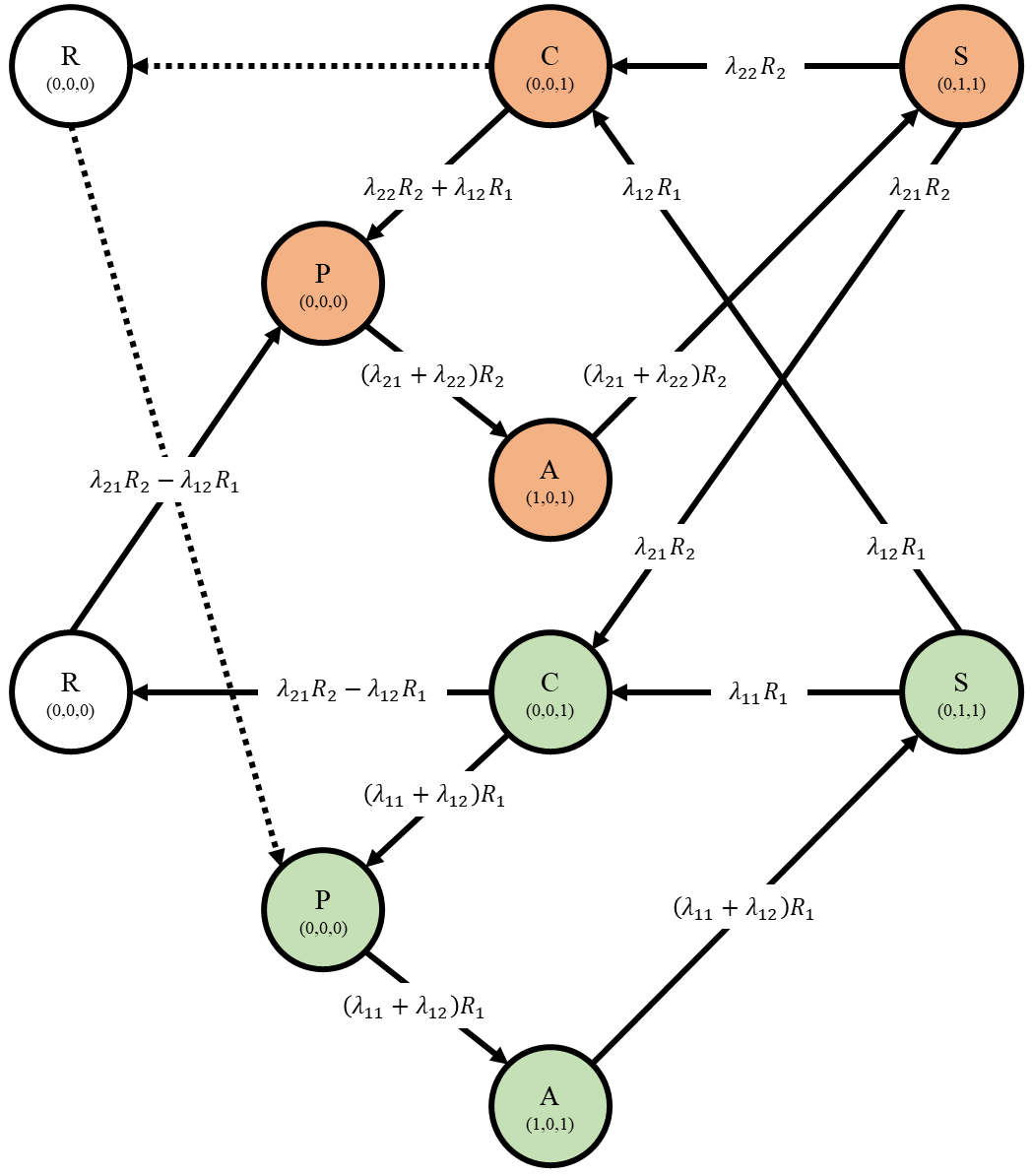}%
      \caption{$\lambda_{12}R_1 < \lambda_{21}R_2$}\label{fig:2112}%
    \end{subfigure}%
    \caption{Graphical Illustration of the average flow in each state transition link in different demand condition}
    \label{fig:twozone_relocation}
\end{figure}


The formulations for deriving the number of SAV fleets in state $R$ are different depending on the demand condition. If there is more passenger demand from zone 1 to zone 2 than there is from zone 2 to zone 1, there will be an increasing number of SAV fleets in zone 2. As a result, some vehicles should be relocated from zone 2 to zone 1. On the other hand, if there is more passenger demand from zone 2 to zone 1 than there is demand from zone 1 to zone 2, there will be an increasing number of SAV fleets in zone 1. In this case, relocation of vehicles should be from zone 1 to zone 2. Figure \ref{fig:twozone_relocation} shows the average flow in each state transition link in both cases.

As a result, if $\lambda_{12}R_1 > \lambda_{21}R_2$,

\begin{equation}
  \begin{split}
  & n_{1,R} ^{req}(t) = 0 \\
  & n_{2,R} ^{req}(t) = (\lambda_{t,12}R_1-\lambda_{t,21}R_2)  T_{R_{21}} = (\lambda_{t,12}R_1-\lambda_{t,21}R_2) \frac{l_{21}}{v_{t,21}} 
  \end{split},
\end{equation}

\noindent
and if $\lambda_{21}R_2 > \lambda_{12}R_1$,

\begin{equation}
  \begin{split}
  & n_{1,R} ^{req}(t)  = (\lambda_{t,21}R_2-\lambda_{t,12}R_1)  T_{R_{12}} = (\lambda_{t,21}R_2-\lambda_{t,12}R_1) \frac{l_{12}}{v_{t,12}} \\
  & n_{2,R} ^{req}(t)  = 0
  \end{split},
\end{equation}

\noindent
where $T_{R_{ij}}$ refers to the average travel time of vehicles that are being relocated from zone $i$ to zone $j$, and $l_{ij}$ refers to the average trip length of vehicles that are being relocated from zone $i$ to zone $j$. 

$n_{1,R}^{req} (t)$ and $n_{2,R}^{req} (t)$ can have different values depending on passenger demand and parameter settings. Consequently, it is not feasible to find a closed-form solution for the optimal values of the operation variables. Thus, we ran a numerical analysis to find the optimal values. We first defined four time windows that maximized and minimized the required fleet size at each zone, as follows:

\begin{equation}
  \begin{split}
  & t_{1,max} = \arg\max_t \left(m_1^{req}(t) \right),\\
  & t_{2,max} = \arg\max_t \left(m_2^{req}(t) \right), \\
  & t_{1,min} = \arg\min_t \left(m_1^{req}(t) \right), \\
  & t_{2,min} = \arg\min_t \left(m_2^{req}(t) \right), \\
  \end{split}
\end{equation}

We assume that the average ground speeds within the same zone in the time windows, $v_{11,{t_{1,max}}}$, $v_{22,{t_{2,max}}}$, $v_{11,{t_{1,min}}}$, and ,$v_{22,{t_{2,min}}}$, are the slowest and fastest of the day in each zone, i.e., $v_{11,{t_{11,max}}}=v_{11,min}$, $v_{22,{t_{22,max}}}=v_{22,min}$, $v_{11,{t_{11,min}}}=v_{11,max}$, and $v_{22,{t_{22,min}}}=v_{22,max}$. Furthermore, we assume that the average speeds between two zones are constant: $v_{t,12} = v_{12}$ and $v_{t,21}=v_{21}$.

Then, we can rewrite the equations of the minimum required fleet size of each zone, $m_1^*$ and $m_2^*$ as follows:

\begin{equation}
  \begin{split}
  & m_1^* = m_1^{req} (t_{1,max}) \\
  & m_2^* = m_2^{req} (t_{2,max}) \\
  \end{split}
\end{equation}

Based on the operation depicted in Figure \ref{fig:twozonescenario}, extra vehicles on the top of required fleet sizes for states $A$, $S$, $C$, and $R$ are not needed. In other words:

\begin{equation}
  \begin{split}
& n_{1,A}(t) = n_{1,A}^{req}(t), n_{1,S}(t) = n_{1,S}^{req}(t), n_{1,C}(t) = n_{1,C}^{req}(t), n_{1,R}(t) = n_{1,R}^{req}(t) \\
& n_{2,A}(t) = n_{2,A}^{req}(t), n_{2,S}(t) = n_{2,S}^{req}(t), n_{2,C}(t) = n_{2,C}^{req}(t), n_{2,R}(t) = n_{2,R}^{req}(t) \\
  \end{split}
\end{equation}

On the other hand, the number of vehicles parked at parking stations in $t$, $n_P (t)$, is not always the same as $n_P^{req} (t)$, but can be found as Equation \ref{eq:tappm_np}:

\begin{equation}
  \begin{split}
& n_{1,P} (t) = m_1^* - \left( n_{1,A}(t) + n_{1,S}(t) + n_{1,C}(t) + n_{1,R}(t) \right) \\  
& n_{2,P} (t) = m_2^* - \left( n_{2,A}(t) + n_{2,S}(t) + n_{2,C}(t) + n_{2,R}(t) \right) \\  
  \end{split}
  \label{eq:tappm_np}
\end{equation}

The number of vehicles not parked in stations in zone $i$, $n_{i,A}(t)+n_{i,S}(t)+n_{i,C}(t)+n_{i,R}(t)$, is the lowest in $t_{i,min}$, so the number of parked vehicles is the highest in $t_{i,min}$. The minimum required number of parking spaces in zone $i$, $y_i R_i$, is the summation of the daily maximum number of required parking spaces, i.e., $n_P (t_{i,min} )$ and additional buffer spaces to guarantee that each parking station is not full by confidence level $q$. As a result, the optimal density of parking spaces can be found as shown in Equation \ref{eq:tappm_y}

\begin{equation}
  \begin{split}
& y^*_1 = \frac{m_1^* - \left( n_{1,A}(t_{1,min}) + n_{1,S}(t_{1,min}) + n_{1,C}(t_{1,min}) + n_{1,R}(t_{1,min}) \right) + \Phi^{-1}(q)\sqrt{2\left(\lambda_{t,11}R_1 + \lambda_{t,21}R_2 \right)HIx_1} }{R_1}\\  
& y^*_2 = \frac{m_2^* - \left( n_{2,A}(t_{2,min}) + n_{2,S}(t_{2,min}) + n_{2,C}(t_{2,min}) + n_{2,R}(t_{1,min}) \right) + \Phi^{-1}(q)\sqrt{2\left(\lambda_{t,22}R_2 + \lambda_{t,12}R_1 \right)HIx_2} }{R_2}\\  
  \end{split}
  \label{eq:tappm_y}
\end{equation}

The density of parking stations ($x_i$) can be calculated similar to Equation \ref{eq:xTa} as shown below:

\begin{equation}
  \begin{split}
  & x_1 = \frac{\kappa^2}{\left(v_{t_{1,max},11}\right)^2} \cdot \frac{1}{\left(T^1_{A,t_{1,max}} \right)^2}\\
  & x_1 = \frac{\kappa^2}{\left(v_{t_{2,max},22}\right)^2} \cdot \frac{1}{\left(T^1_{A,t_{2,max}} \right)^2}\\
  \end{split},
\end{equation}

The objective function in Equation \ref{eq:ob_tappm} can be reformulated in terms of $T^1_{A,t_{1,max}}$ and $T^1_{A,t_{2,max}}$. We numerically find the optimal values for $T^1_{A,t_{1,max}}$ and $T^1_{A,t_{2,max}}$ that minimizes total operation cost.

\clearpage
\section{Case Study for Two-Zone Analytical Parking Planning Model}\label{sec:case_study_tappm}


In this section, we will extend the case study in Section \ref{sec:case_study_sappm} and discuss the findings for T-APPM. The main difference between S-APPM and T-APPM is that T-APPM considers inter-region passenger demand as well as the relocation of the SAV fleet. We extend the spatial range to the Seoul Metropolitan Area (i.e. Seoul Capital Area, or Sudogwon).

\begin{figure}[!t]
  \centering
  \includegraphics[width=0.7\textwidth]{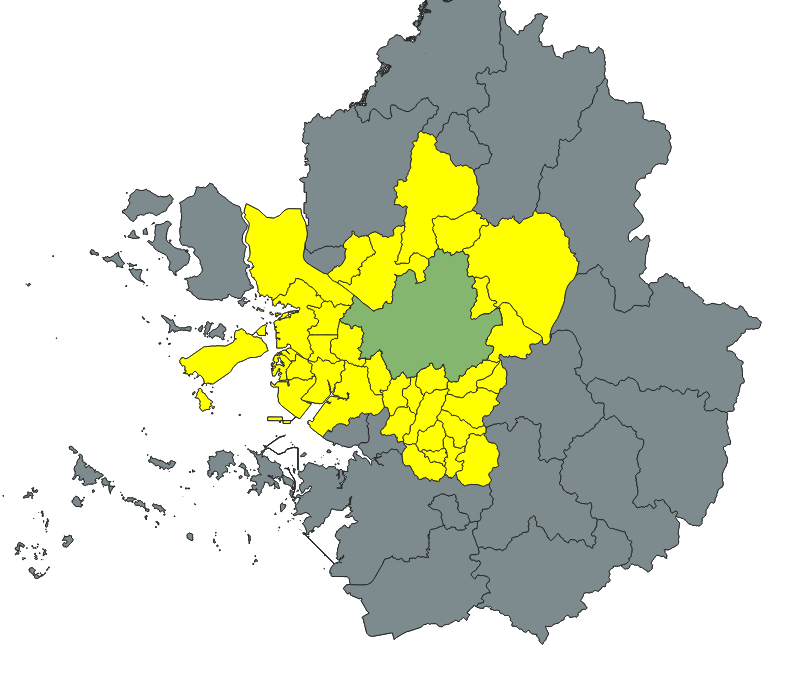}
  \caption{Map of Seoul Metropolitan Area. The green region represents Seoul city, while the yellow and gray regions represent Gyeonggi-do and Incheon city. Yellow regions are selected as the target region in Gyeonggi-do in this study. }\label{fig:SMA}
\end{figure}


The Seoul Metropolitan Area refers to the metropolitan area near Seoul, including Seoul, Incheon, and Gyeonggi-do, located in the northwest part of South Korea. The population of this area is approximately 26 million people, more than half the population of South Korea. Figure \ref{fig:SMA} provides a map of the Seoul Metropolitan Area. The green region is Seoul city, while the yellow and gray regions are Gyeonggi-do and Incheon city. Gyeonggi-do and Incheon city cover a wide range of regions, as shown in Figure \ref{fig:SMA}. As a result, if we consider the whole area as the target region for our case study, intra-region travel times and inter-region travel times will be too long. It would be inefficient and unrealistic for SAVs to travel long distances for relocation and passenger trips. Therefore, we select sub-regions near Seoul that have a relatively considerable number of passenger trips to Seoul. The selected area is colored in yellow in Figure \ref{fig:SMA}; this region will be referred to as \textit{Gyeonggi} for the rest of this case study.


Table 6 shows the average hourly unit passenger demand for our case study according to the Korean National Household Travel Survey (Origin-Destination Flow Survey). Similar to Table 3, the table shows corresponding values for each time window. Unit passenger demand is calculated based on origin. For example, if there were 52,982.71 passengers traveling from Seoul to Gyeonggi, we divided the number of passengers by the area of the origin zone, and the result was 87.54 [$veh/km^2/hr$]. Similar to Table 3, the values in the total row represent average passenger demand during any time-of-day. The values in the \textit{AM peak} row represent average passenger demand during the morning peak (7-9 AM), and the values in the \textit{PM peak} row represent average passenger demand during the afternoon peak (6-8 PM). In this study, we assume that passenger demand values in time windows other than AM and PM peaks are equal to off-peak passenger demand. As a result, the values in the \textit{Off-peak} row were calculated based on the values in the Total, AM peak, and PM peak rows.

\begin{table}[H]
	\caption{Hourly average passenger demand in Seoul. The unit of the values is $[veh/km^2/hr]$}
	\begin{center}
	\begin{tabular}{c|l|l|c|c}
	     Time & Origin & Destination & \makecell{Passenger Demand \\ by Personal Vehicle} & \makecell{Passenger Demand \\ by All Mode} \\\hline\hline 
	     \multirow{4}{*}{Total} 
	    & Seoul & Seoul & 285.11 & 1720.03 \\\cline{2-5}
        & Seoul & Gyeonggi & 87.54 & 191.06 \\\cline{2-5}
        & Gyeonggi & Seoul & 18.90 & 40.80 \\\cline{2-5}
        & Gyeonggi & Gyeonggi & 79.33 & 232.69 \\\hline\hline
	     \multirowcell{4}{AM peak \\ (7-9 AM)} 
	     & Seoul & Seoul & 765.04 & 4518.16 \\\cline{2-5}
        & Seoul & Gyeonggi & 280.37 & 448.92 \\\cline{2-5}
        & Gyeonggi & Seoul & 70.32 & 165.51 \\\cline{2-5}
        & Gyeonggi & Gyeonggi & 216.07 & 619.37 \\\hline\hline
	     \multirowcell{4}{PM peak \\ (6-8 PM) }  
	     & Seoul & Seoul & 836.94 & 4042.69 \\\cline{2-5}
        & Seoul & Gyeonggi & 340.25 & 795.70 \\\cline{2-5}
        & Gyeonggi & Seoul & 58.01 & 85.71 \\\cline{2-5}
        & Gyeonggi & Gyeonggi & 225.83 & 525.61 \\\hline\hline
	     \multirow{4}{*}{Off peak} 
	     & Seoul & Seoul & 181.93 & 1207.95 \\\cline{2-5}
        & Seoul & Gyeonggi & 42.98 & 104.81\\\cline{2-5}
        & Gyeonggi & Seoul & 9.85 & 23.84 \\\cline{2-5}
        & Gyeonggi & Seoul & 51.01 & 164.73 \\
	     \hline\hline
	\end{tabular}
	\end{center}
	\label{tab:sg_demand}
\end{table}

\begin{table}[H]
	\caption{Model parameters used for the case study.}
	\begin{center}
	\begin{tabular}{l|l|l|l}
	\multicolumn{2}{l|}{Variable}  & Units & Value \\ \hline\hline
	\multirow{2}{*}{$R$}
	& $R_s$ & $\mathrm{[km^2]}$  & 605.24 \\ \cline{2-4}
	& $R_g$  & $\mathrm{[km^2]}$ & 2799.20 \\ \hline
	\multirow{4}{*}{$L$}
	& $l_{ss}$ & $\mathrm{[km]}$  & 14.76 \\ \cline{2-4}
	& $l_{sg}$ & $\mathrm{[km]}$ & 25.48 \\\cline{2-4}
	& $l_{gs}$ & $\mathrm{[km]}$& 25.48 \\\cline{2-4}
	& $l_{gg}$ & $\mathrm{[km]}$& 31.74 \\\hline
	\multirow{4}{*}{$V$}
	& $v_{ss,min}$, $v_{ss,max}$ & $\mathrm{[km/hr]}$& 18.0, 40.0  \\\cline{2-4}
	& $v_{sg,min}$, $v_{sg,max}$ & $\mathrm{[km/hr]}$& 25.0, 35.0  \\\cline{2-4}
	& $v_{gs,min}$, $v_{gs,max}$ & $\mathrm{[km/hr]}$& 25.0, 35.0  \\\cline{2-4}
	& $v_{gg,min}$, $v_{gg,max}$ & $\mathrm{[km/hr]}$& 20.0, 50.0  \\\hline
    \multicolumn{2}{l|}{$p$}   & - & 0.95 \\ \hline
    \multicolumn{2}{l|}{$q$}  & - & 0.95 \\ \hline
    \multicolumn{2}{l|}{$\alpha$} & - & 2 \\ \hline
    \multicolumn{2}{l|}{$I$} & -  & 1 \\ \hline
    \multicolumn{2}{l|}{$H$}  & -  & 2\\ \hline
    \multicolumn{2}{l|}{$\kappa$}  & - & 0.5\\ \hline
    \multicolumn{2}{l|}{$T_0$}  & [$\mathrm{hr}$] & 1/60 \\ \hline\hline
    \multirow{4}{*}{$C$}
    & $C_m$ & $\mathrm{[\$/veh/day]}$ & 35.616 \\\cline{2-4}
    & $C_x$& $\mathrm{[\$/stations/day]}$  & 1 \\\cline{2-4}
    & $C_{y,s}$  & $\mathrm{[\$/spaces/day]}$ & 4.73 \\\cline{2-4}
    & $C_{y,g}$ & $\mathrm{[\$/spaces/day]}$ & 0.24 \\ \hline\hline
	\end{tabular}
	\end{center}
	\label{tab:tappm_parameters}
\end{table}

\begin{table}[H]
	\caption{Summary of results of Case Study for T-APPM}
	\label{tab:twozone_result_summary}
	\begin{center}
	\begin{tabular}{l|ll|l|l|l|l}
  Demand Type & Variable & & Current &\makecell{ Seoul Only \\ S-APPM} & \makecell{Gyeonggi Only \\ S-APPM} & T-APPM  \\\hline\hline
  \multirowcell{9}{Passenger Demand \\ by Personal Vehicle}
 & $x_{s}$ &$\mathrm{[stations/km^2]}$& 524.09 & 11.66 & -  & 13.36  \\\cline{2-7}
 & $x_{g}$ &$\mathrm{[stations/km^2]}$& 97.92 & - & 10.32 & 8.16 \\\cline{2-7}
 & $y_{s}$ &$\mathrm{[spaces/km^2]}$& 7,150.72 & 718.22 & -  & 820.12   \\\cline{2-7}
 & $y_{g}$ &$\mathrm{[spaces/km^2]}$& 1740.12 & - & 175.66 & 366.54  \\\cline{2-7}
 & $z_{s}$ &$\mathrm{[spaces/station]}$& 13.64 & 61.59 & - & 61.39  \\\cline{2-7}
 & $z_{g}$ &$\mathrm{[spaces/station]}$& 17.77 & - & 17.01 & 44.90  \\\cline{2-7}
 & $m_{s}$ &$\mathrm{[veh]}$& 2,703,429 & 477,944 & -  & 605,699 \\\cline{2-7}
 & $m_{g}$&$\mathrm{[veh]}$& 4,394,130 & - & 536,827  & 1,327,406 \\\cline{2-7}
 & $m$ & $\mathrm{[veh]}$ & 7,097,559 & - & - & 1,933,105  \\\hline\hline
   \multirowcell{9}{Passenger Demand \\ by All Mode}
 & $x_{s}$ &$\mathrm{[stations/km^2]}$& 524.09 & 27.512 & -  & 28.54  \\\cline{2-7}
 & $x_{g}$ &$\mathrm{[stations/km^2]}$& 97.92 & - & 17.68 & 13.12  \\\cline{2-7}
 & $y_{s}$ &$\mathrm{[spaces/km^2]}$& 7,150.72 & 3728.66 & -  & 3758.08   \\\cline{2-7}
 & $y_{g}$ &$\mathrm{[spaces/km^2]}$& 1740.12 & - & 469.24 & 640.92  \\\cline{2-7}
 & $z_{s}$ &$\mathrm{[spaces/station]}$& 13.64 & 135.52 & - & 131.68  \\\cline{2-7}
 & $z_{g}$ &$\mathrm{[spaces/station]}$& 17.77 & - & 26.53 & 48.85  \\\cline{2-7}
 & $m_{s}$ &$\mathrm{[veh]}$& 2,703,429 & 2,549,647 & -  & 2,583,452 \\\cline{2-7}
 & $m_{g}$ &$\mathrm{[veh]}$& 4,394,130 & - & 1,460,981  & 2,344,356 \\\cline{2-7}
 & $m$ &$\mathrm{[veh]}$& 7,097,559 & - & - & 5,123,078  \\\hline
	     \hline	     
	\end{tabular}
	\end{center}
\end{table}

Table \ref{tab:tappm_parameters} shows the model parameters used for the case study. The area of Seoul is 605.24 $\mathrm{km^2}$; the area of Gyeonggi (selected areas only) is 2799.20 $\mathrm{km^2}$. $l_{ij}$ represents the average travel distance from zone $i$ to zone $j$. The passenger demand is assumed to be uniformly distributed, so $l_{ij}$ is the average distance between one random point in zone $i$ and one random point in zone $j$. We use the following approximation from \cite{rodriguez1983errors} and \cite{wilson1990average} to calculate $l_{ij}$:


\begin{equation}\label{eq:l_approx}
  \begin{split}
  l_{ij} 
  & = \sqrt{\left(\left(\Bar{d}_{i*}\right)^2 + \left(\Bar{d}_{j*}\right)^2\right) + \left(\Bar{d}^*_{ij}\right)^2} \\
  & \approx \sqrt{0.18\left(R_i + R_j\right) + \left(\Bar{d}^*_{ij}\right)^2}\\
  \end{split},
\end{equation}

\noindent
where $\Bar{d}_{i*}$ is the average distance from one point in zone $i$ to the centroid of zone $i$, and $\Bar{d}^*_{ij}$ is the distance between the centroids of zone $i$ and zone $j$. In the Table \ref{tab:tappm_parameters}, the subscript $s$ refers to Seoul and the subscript $g$ refers to Gyeonggi. For example, $l_{sg}$ refers to the average travel distance from Seoul to Gyeonggi.
%


Because we covered the sensitivity analysis of different cost values in Section 3, we assumed in this section that the costs are pre-determined. We used 35.36 $\$/veh/day$ for $C_m$ referenced from \cite{estrada2021operational}.$C_{y,1}$ is assumed to be 4.73 $\$/spaces/day$ and $C_{y,2}$ is assumed to be 0.24 $\$/spaces/day$, which are the amortized land cost of two zones respectively, referenced from declared land value announced by the Korean Ministry of Land, Infrastructure, and Transport. We assumed that $C_x$ is 1 $\$/stations/day$ to make sure that C x was neither too large nor too small compared to the other cost variables.

The results of this case study are shown in Table \ref{tab:twozone_result_summary}. To analyze the effect of relocation and inter-zonal demands in T-APPM, we compare the results of T-APPM with  three baselines. The first baseline is the current operating values in Seoul and Gyeonggi, denoted as ``Current'' in Table \ref{tab:twozone_result_summary}. The second and third baselines are the results of S-APPM introduced in Section \ref{SAPPM}, denoted as ``Seoul Only S-APPM'' and ``Gyeonggi Only S-APPM.'' ``Seoul Only S-APPM'' refers to the results of S-APPM when only the intra-zonal passenger demand in Seoul is considered as in Section \ref{sec:case_study_sappm}, and ``Gyeonggi Only S-APPM'' refers to the results of S-APPM when only the intra-zonal passenger demand in Gyeonggi is considered.

In Table \ref{tab:twozone_result_summary}, it can be seen that $x_s$ increased while $x_g$ decreased in both demand scenarios. Since $x$ is proportional to the squared reciprocal of $T_{A,t}^1$, this result implies that the average passenger waiting time in Seoul decreased and the average passenger waiting time in Gyeonggi increased when the two zones were considered together. The results of $y$ are notable. In both demand scenarios, the density of parking spaces in Seoul ($y_s$) increased slightly, while the density of parking spaces in Gyeonggi ($y_g$) increased significantly. This shows that increasing the number of parking spaces in Gyeonggi is more cost-efficient than increasing the number of parking spaces in Seoul. These results show that T-APPM is capable of incorporating different cost variables across two regions so that the most cost-efficient solution can be derived. The densities of parking spaces in both Seoul and Gyeonggi are significantly less than the ``Current'' values of Seoul and Gyeonggi. This result implies that if the SAV system is introduced, the number of parking spaces will be significantly reduced. As a result of changes in x and y, the average number of parking spaces at each parking station in Seoul $z_s$ changed slightly, while that of Gyeonggi $z_g$ increased significantly. From this result, it can be concluded that, with the SAV system, it is more cost-efficient to have fewer parking stations with greater numbers of parking spaces, and especially to install more parking spaces in suburbs, which have lower land costs, reducing overall cost.
Both $m_s$ and $m_g$ increase in T-APPM results because of increases in passenger demand. However, the results show that, compared to the ``Current'' situation, it is possible to significantly decrease the total number of vehicles by using the SAV system. Even when SAV replaces all passenger modes, including train, bus, and subway, and serves as a primary mode of passenger trips (i.e., the result of Passenger Demand by All-Mode, shown in Table \ref{tab:twozone_result_summary}), the total number of vehicles is still less than the “Current” value, which is currently only used for passenger demand by personal vehicles.
%





\clearpage
\section{Conclusion}\label{sec:conclusion}
This study presents two analytical models to describe parking operations for SAVs in a given urban transportation system: the Single-zone Analytical Parking Planning Model (S-APPM) and the Two-zone Analytical Parking Planning Model (T-APPM). S-APPM assumes that the given traffic network is a single network with homogeneous network characteristics and passenger demand. On the other hand, T-APPM assumes that the given traffic network contains two separate zones, usually the city center and suburb. Both models are carefully derived based on the general model of demand-responsive transportation services \citep{daganzo2019general}, introducing novel concepts such as parking and relocation. S-APPM offers a closed-form solution for parking operation scenarios with single-zone intra-zonal passenger trips. Consequently, computational complexity is significantly reduced compared to previously studied simulation-based methodologies. Extending S-APPM, T-APPM introduces inter-zonal passenger trips and the relocation of vehicles. The solution of the case study for T-APPM shows that this model can incorporate different macroscopic characteristics across two zones.

The contribution of this study is that two models allow policy-makers and decision-makers to plan parking operations under the dominance of Shared Autonomous Vehicles, yielding approximated numbers for both densities of parking stations and parking spaces. Using the proposed models to find optimal operational variables is much simpler than using previously studied simulation-based approaches. The simulation-based methods require much effort and time to set the simulation environment and run the simulation with different variables. However, the solutions from the proposed models can be derived in much less time and have much less complexity.



There are several assumptions made during the model derivations; these assumptions can be further studied so that the model can be made more realistic. As discussed in Section \ref{section:2.1.2}, it is possible to allocate a new passenger request to an SAV cruising back to the parking station (state C). The dotted line in Figure \ref{fig:modelstate} indicates this state transition. Introducing this state transition would improve the efficiency of the whole system and, as a result, results will be better with a smaller SAV fleet and fewer parking spaces. Also, it is assumed in T-APPM that SAVs are relocated to other zones only from state C. However, it is also realistic to assume that relocation of vehicles can happen from state P. This change can improve the model by dealing with cases in which extreme relocation of vehicles is required, and vehicles parked in one zone must be relocated to the other zone. Finally, we used several approximations in both case studies in Section \ref{sec:case_study_sappm} and Section \ref{sec:case_study_tappm}. The results of the case studies can be more realistic if we use field-observed values instead of approximated values. Using urban vehicle trajectory data is desirable to achieve adequate values for each parameters \citep{naveh2018urban,choi2021trajgail,jin2022transformer}.


Future research can be conducted in various directions. In T-APPM, for simplicity of the model, we considered only inter-zonal passenger trips between two zones. However, T-APPM can be further investigated using a multi-zone approach. For example, we assumed in Sections \ref{sec:case_study_sappm} and \ref{sec:case_study_tappm} that the macroscopic characteristics of Seoul and Gyeonggi, such as passenger demand and costs, were homogeneous. However, Seoul and Gyeonggi can be split into multiple zones with different macroscopic characteristics. To create a ``multi-zonal'' analytical parking planning model, it is necessary to consider much more complex relocation between zone pairs. Such improvements will answer questions like `Where exactly should parking stations be located?' In addition, since most SAVs are likely to be electric vehicles (EV), it is also necessary to consider battery and charging infrastructure \citep{lee2021optimal} when planning the parking operation. In the incoming era of SAVs, parking stations will serve as storage places for unused vehicles and as depots that manage overall SAV operation, including charging. As a result, incorporating existing studies on the battery cycle of EVs and the charging infrastructure into parking operation is a good direction for future research.

%
%
%

\clearpage

\clearpage

\printcredits

\bibliographystyle{cas-model2-names}

\bibliography{cas-refs}

\clearpage
\appendix
\section{Glossary of Notations}\label{tab:notations}

\restylefloat{table}
\begin{table}[h]
	\begin{center}
	\begin{tabular}{p{1.5cm}|p{2.5cm}|p{10cm}}
	     Notation & Unit & Meaning  \\\hline\hline
	     $x$ & $\mathrm{[stations/km^2]}$ & density of parking stations \\\hline
	     $y$ & $\mathrm{[spaces/km^2]}$ & density of parking spaces \\\hline
	     $m$ & $\mathrm{[veh}]$ & number of SAV fleets \\\hline
         $z$ & $\mathrm{[spaces/stations]}$ & the average number of parking spaces in a parking station \\\hline
	     $R$ & $\mathrm{[km^2]}$ & area of target area \\\hline
	     $Cost$ & $\mathrm{[\$/day]}$ & overall daily average operation cost \\\hline
	     $T_{A,t}$ & $\mathrm{[hr]}$ & average passenger waiting time \\\hline
	     $T^i_{A,t}$ & $\mathrm{[hr]}$ & average travel time from the $i$-th nearest parking station to the origin of the passenger \\\hline
	     $T_{0}$ & $\mathrm{[hr]}$ & maximum allowed average passenger waiting time \\\hline
	     $C_x$ & $\mathrm{[\$/stations/day]}$ & daily operation cost of a parking station \\\hline
	     $C_y$ & $\mathrm{[\$/spaces/day]}$ & daily operation cost of a parking space \\\hline
	     $C_m$ & $\mathrm{[\$/vehicles/day]}$ &  daily operation cost of an SAV fleet \\\hline
	     $n_A$ & $\mathrm{[veh]}$ & the number of SAV fleets in state $A$ \\\hline
	     $n_S$ &  $\mathrm{[veh]}$ & the number of SAV fleets in state $S$ \\\hline
	     $n_C$ &  $\mathrm{[veh]}$ & the number of SAV fleets in state $C$ \\\hline
	     $n_P$ &  $\mathrm{[veh]}$ & the number of SAV fleets in state $P$ \\\hline
	     $n_A^{req}$ &  $\mathrm{[veh]}$ & the required number of SAV fleets in state $A$ \\\hline
	     $n_S^{req}$ &  $\mathrm{[veh]}$ & the required number of SAV fleets in state $S$ \\\hline
	     $n_C^{req}$ &  $\mathrm{[veh]}$ & the required number of SAV fleets in state $C$ \\\hline
	     $n_P^{req}$ &  $\mathrm{[veh]}$ & the required number of SAV fleets in state $P$ \\\hline
	     $v_{min}$&  $\mathrm{[veh]}$ & the minimum average ground speed throughout the day \\\hline
	     $v_{max}$&  $\mathrm{[veh]}$ & the maximum average ground speed throughout the day \\\hline
	     $\lambda_t$ & $\mathrm{[veh/hr/km^2]}$ & average passenger demand in time window $t$ \\\hline
	     $l_t$ & $\mathrm{[km]}$ & average trip length \\\hline
	     $v_t$ & $\mathrm{[km/hr]}$ & average ground speed in time window $t$ \\\hline
	     $T_{S,t}$ & $\mathrm{[veh]}$ & average travel time for passenger trip\\\hline
	     $T_{C,t}$ & $\mathrm{[veh]}$ & average travel time from the destination of the passenger to the nearest parking station \\\hline
	     $T^i_{C,t}$ & $\mathrm{[veh]}$ & average travel time from the destination of the passenger to the $i$-th nearest parking station \\\hline
	     $p$ & - &  confidence level to guarantee that the passengers are assigned to the nearest parking station at a certain confidence level.\\\hline
	     $q$ & - & confidence level guarantee that the nearest parking station is not full.\\\hline
	     $\alpha$ & - & the incremental ratio of the travel time to the next nearest parking station \\\hline	     
	     $H$ & $\mathrm{[hr]}$ & length of a time window \\\hline
	     $I$ & - & the mean-to-variance ratio of the number of fleets parked at each parking station \\\hline
	     \hline
	\end{tabular}
	\end{center}
\end{table}

\end{document}